\documentclass{emulateapj}
\usepackage{graphicx,amsfonts,natbib,longtable,apjfonts,lscape}
\shorttitle{Ammonia in Perseus}
\shortauthors{Rosolowsky et al.}
\begin{document}
\title{An Ammonia Spectral Atlas of Dense Cores in Perseus}
\author{E. W. Rosolowsky\altaffilmark{1,2}, J. E. Pineda\altaffilmark{2}, J.  B. Foster\altaffilmark{2},
  M. A. Borkin\altaffilmark{3}, J. Kauffmann\altaffilmark{2,3}, P. Caselli\altaffilmark{2,4}, P. C. Myers\altaffilmark{2}
  and A. A. Goodman\altaffilmark{2,3}} 
 \email{erosolow@cfa.harvard.edu}

\altaffiltext{1}{National Science Foundation Astronomy and
  Astrophysics Postdoctoral Fellow}

\altaffiltext{2}{Center for Astrophysics, 60 Garden St, Cambridge,
  MA 02138}

\altaffiltext{3}{Initiative for Innovative Computing, Harvard
  University, 60 Oxford St., Cambridge, MA 02138}

\altaffiltext{4}{School of Physics and Astronomy,
University of Leeds, Leeds LS2 9JT, UK}

\begin{abstract}
We present ammonia observations of 193 dense cores and core candidates
in the Perseus molecular cloud made using the Robert F. Byrd Green
Bank Telescope.  We simultaneously observed the NH$_3$(1,1),
NH$_3$(2,2), C$_2$S ($2_1\to 1_0$) and C$_2 ^{34}$S($2_1\to 1_0$)
transitions near $\nu = 23$ GHz for each of the targets with a
spectral resolution of $\delta v \approx 0.024 \mbox{ km s}^{-1}$.  We
find ammonia emission associated with nearly all of the
(sub)millimeter sources as well as at several positions with no
associated continuum emission.  For each detection, we have
measured physical properties by fitting a simple model to every
spectral line simultaneously.  Where appropriate, we have refined the
model by accounting for low optical depths, multiple components along
the line of sight and imperfect coupling to the GBT beam.  For the
cores in Perseus, we find a typical kinetic temperature of $T_k=11$~K,
a typical column density of $N_{\mathrm{NH3}}\approx 10^{14.5}\mbox{
  cm}^{-2}$ and velocity dispersions ranging from $\sigma_v =
0.07\mbox{ km s}^{-1}$ to 0.7$\mbox{ km s}^{-1}$.  However, many cores
with $\sigma_v >0.2\mbox{ km s}^{-1}$ show evidence for multiple
velocity components along the line of sight.
\end{abstract}

\keywords{ISM:clouds --- ISM:molecules --- radio lines:ISM}

\section{Introduction}
Ammonia remains one of the best molecules for studying the cool, dense
molecular cores where most stars form.  The utility of ammonia was
recognized early in the pursuit of molecular line astronomy \citep[and
  references therein]{nh3-araa} and it remains the standard for
identifying and studying the internal conditions of dense molecular
cores. The unique quantum structure of the molecule coupled with its
relative abundance allows for a host of measurements to be made from
the hyperfine transitions among the multiple metastable states, which
emit near $\nu=23$ GHz.  A single spectrum can be used to determine
the line-of-sight velocity, velocity dispersion and gas kinetic
temperature for a dense core.  This set of properties form an
excellent complement to surveys of submillimeter emission
\citep[e.g.][]{motte-andre,testi-sargent} which readily study the size
and distribution of the dust emission in cores, while yielding no
information about the kinematics and temperatures.

Recently, submillimeter surveys of dense cores have been extended to
cover large fractions of molecular clouds
\citep{hatchell05,bolocam-perseus}, making complete surveys of dense
cores possible.  In addition, large scale mapping projects have
surveyed several nearby molecular clouds at high resolution in
emission from the isotopomers of CO \citep[e.g., the COMPLETE Surveys
  of Serpens, Ophiuchus, Perseus;][]{complete-data}.  The CO surveys
establish the cores in the larger context of the molecular cloud.
However, these surveys have raised many questions about the properties
of cores and their relationship to the larger molecular environment.

To measure the kinetic temperature and kinematics of an unbiased
sample of dense cores embedded in the same molecular complex, we have
conducted a survey of dense cores in ammonia across the Perseus
molecular cloud using the Green Bank Telescope.  There are several
advantages to adopting Perseus as a target.  It has been extensively
studied in several observational campaigns: the COMPLETE survey of
star forming regions which surveyed the molecular gas using the FCRAO
14-m \citep{complete-data}; the SCUBA survey of submillimeter emission
\citep{hatchell05,perseus-scuba}; the BOLOCAM survey of the region in
the 1.1 mm continuum \citep{bolocam-perseus}; and in all the Spitzer
bands by the c2d project \citep{c2d-pers-irac,c2d-pers-mips}.  The
locations of dense cores have been identified in the (sub)millimeter
maps, and their protostellar content has been explored
\citep{jorgensen-cores,hatchell-cores}.  In addition, Perseus shows a
wide range of star forming environments, ranging from the newly formed
clusters IC 348 and NGC 1333 to more isolated star forming regions
such as B5 and L1448.  A substantial portion of the molecular mass in
the cloud is not currently forming stars.

Several previous observational studies provide context for the
observations of Perseus.  The observational results are homogenized
in \citet{jijina}.  The typical cores in Perseus have
$\log(\mathrm{NH}_3/\mbox{cm}^{-2})=14.5$, $R_{pc}=0.09~\mbox{ pc}$
(after rescaling to our preferred distance of 260 pc), velocity
dispersion $\sigma_v=0.17\mbox{ km s}^{-1}$, $T_{k}=11~\mbox{K}$.
However, these studies have primarily observed the well known star
forming regions with less concern for objects in the sterile portions
of the molecular cloud.  

This survey presents observations of NH$_3$ and C$_2$S emission from a
variety of sources in Perseus including millimeter-bright dense cores
as well as otherwise unremarkable high column density features
selected from far infrared emission.  The two tracers present
complimentary views of the chemical evolution of the cloud.  C$_2$S is
regarded as an ``early-time'' tracer formed in the initial conversion
of atomic to molecular gas and excited at high densities
\citep[$n_{cr}\sim 10^{4.5}\mbox{ cm}^{-3}$][]{langer95,pp5-difran}.
As carbon species are depleted, C$_2$S disappears.  In contrast,
NH$_3$ is regarded as a late-time tracer like N$_2\mbox{H}^+$, with
both species requiring the relatively slow formation of N$_2$ as a
precursor.  The molecules are excited at similar densities as C$_2$S
but should not appear in significant amounts until later in the
protostellar collapse \citep[$t\sim 10^{5.5}\mbox{
yr}$][]{flower06,pp5-difran}.  \citet{tafalla04} note that the
abundance of NH$_3$ varies by a factor of several across starless
cores while N$_2\mbox{H}^+$ remains constant.  Such variations can
complicate using ammonia as a structural tracer, but the utility of
having a direct temperature measurement from the NH$_3$ is immense.

Our survey of dense cores in Perseus adopted the form of a spectral
survey to maximize the number of cores we could sample in a limited
amount of time.  Even with a single spectrum, we are able to determine
core kinematics, velocity dispersion, kinetic temperatures, and
chemical abundances of ammonia and C$_2$S.  We also spent a
significant amount of time surveying core candidates derived through a
variety of methods to find ammonia emission from objects not bright in
the submillimeter.  In this paper, we present the results of our
survey and derive physical parameters from the ammonia spectra.  A
detailed comparison of the core properties to other tracers will be
presented elsewhere.

\section{Observations}
\label{observations}
We observed 193 dense cores and core candidates in the Perseus
Molecular Cloud using the 100-m Robert F. Byrd Green Bank Telescope
(GBT).  The observations were conducted from 2 October -- 10 November
2006 in eight separate observing shifts spanning a total of 59 hours.
For each target we conducted single-pointing, frequency-switched
observations for 5-30 minutes depending on the source.  We used the
high-frequency $K$-band receiver and configured the spectrometer to
observe 4 12.5-MHz windows centered on the rest frequencies of
NH$_3$(1,1) \citep[23.6944955(1) GHz,][]{lovas-lines}, NH$_3$(2,2)
\citep[23.7226333(1) GHz,][]{lovas-lines}, CCS ($2_1\to 1_0$)
\citep[22.344033(1) GHz,][]{ccs-freq} and CC$^{34}$S ($2_1\to 1_0$)
\citep[21.930476(1) GHz,][]{cc34s-freq}.  The frequency uncertainties
translate to errors of 1-10 m~s$^{-1}$ uncertainties in our velocity
scale and, for high signal-to-noise lines, limit the accuracy to which
we can centroid the velocity.  The spectrometer produces 8192 lags
across each window yielding 1.525 kHz channel separation with 1.862
kHz resolution (0.024 km~s$^{-1}$ at this frequency) since the lags in
the spectrometer are uniformly weighted.  The frequency switch was
asymmetric with a shift of $\pm 2.0599365$~MHz around the center of
the band, allowing the entire NH$_3$(1,1) complex to remain within the
spectral window.

We updated the pointing model of the telescope with observations of
the quasar 0336+3218 every 45-90 minutes, depending on the wind
conditions.  In nearly all instances, the corrections to the model
were $<10''$ except in the worst wind conditions \citep[the GBT beam
at 23 GHz is $31''$ or 0.04 pc at the assumed 260 pc of
Perseus;][]{cernis93}.  Since the typical dense core size is $\sim
0.08$ pc for Perseus \citep{jijina} pointing deviations should not
confuse sources with the exception of the most densely clustered
regions (IC 348, NGC 1333).  Some of the complex velocity structure
seen in the ammonia spectra almost certainly results from multiple
sources along the line of sight (\S\ref{multicomp}).  However, sources
are chosen to be separated by $\ge 1$ GBT beam FWHM confusion due to
overlapping beams should be negligible compared to confusion intrinsic
to the sources on the sky.

We calibrated the data with injection of a noise signal periodically
throughout the observations.  Because of slow variations in the power
output of the noise diodes and their coupling to the signal path, we
measured the strength of the noise signal through observations of a
source with known flux (the NRAO flux calibrator 3C84).  We repeated
the flux calibration observations during every observing run to detect
any changes in the calibration sources, finding no significant
variations over the course of our run.  Calibrating the noise diodes
established the $T_A$ scale, and we scaled to the $T_A^*$ scale using
estimates of the atmospheric opacity at 22-23 GHz from models of the
atmosphere derived using weather
data\footnote{\url{http://www.gb.nrao.edu/$\sim$rmaddale/Weather/index.html}}.
To reach the $T_{mb}$ scale, the spectra are divided by the main beam
efficiency of the GBT, which is $\eta_{mb}\approx 0.81$ at these
frequencies.  We observed one of our sources (NH3SRC 47, where NH3SRC
is our source catalog designator) every night for 5 minutes and find
$\lesssim 5\%$ changes in the signal amplitude over the course of the
project.  The changes likely result from pointing offsets and
inaccuracies in the opacity model.  The accuracy of our absolute
calibration will affect some of our parameter estimates (such as
column density) in excess of our derived uncertainties
(\S\ref{errors}).  The relative calibration within a spectrum appears
to be better than the noise level in all the spectra of NH3SRC 47.

We subtract a linear baseline from each of the spectra, restricting to
windows outside the expected range for ammonia emission from Perseus
($v_{LSR} = -2 \to 12\mbox{ km s}^{-1}$), including the splitting from
the hyperfine structure.  The velocity window comes from the COMPLETE
observations of $^{13}$CO towards the Perseus cloud
\citep{complete-data}.

\begin{figure*}
\plotone{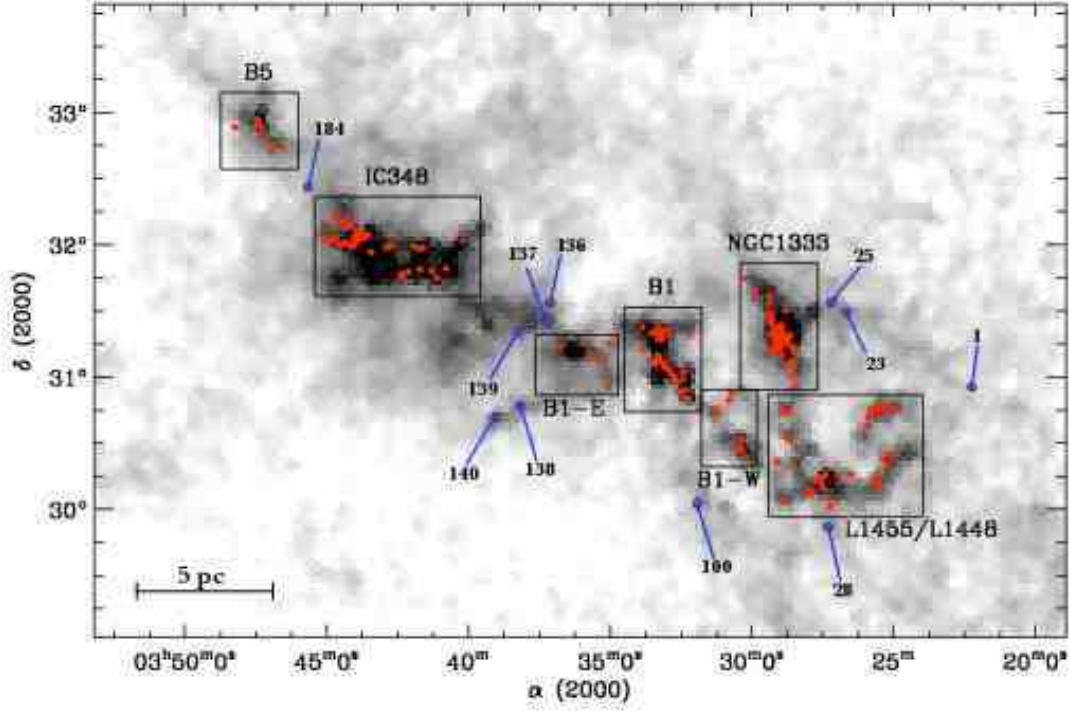}
\caption{\label{findchart1} Locations of GBT pointings (red points)
  overlaid upon a map of extinction for the Perseus region.  Seven
  subregions have been defined and have their sources labeled in
  Figure \ref{findchart2}.  For sources outside the defined
  subregions, the sources are labeled on the map with the number of
  the source given in Table \ref{obstable}.  The extinction map is
  derived from applying the NICER algorithm to 2MASS data
  \citep{complete-data} and the grayscale has a square-root transfer
  function running from $A_V=0$ to 7 mag.}
\end{figure*}

\begin{figure*}
\plotone{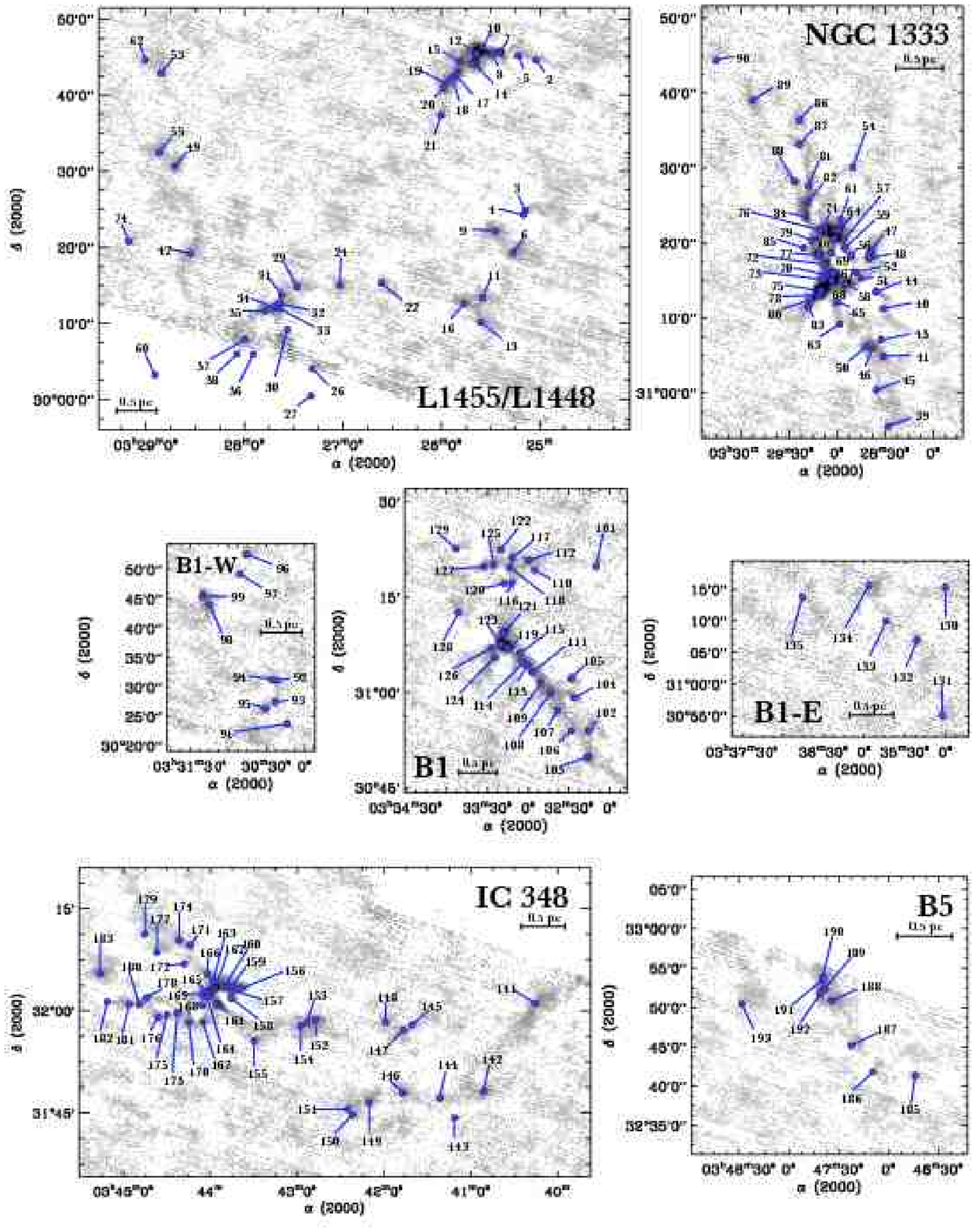}
\caption{\label{findchart2} Locations of GBT pointings (red points)
  overlaid upon a maps of millimeter emission from the BOLOCAM survey
  of the region \citep{bolocam-perseus}. The seven maps correspond to
  the subregions defined in Figure \ref{findchart1}.  Individual
  pointings are labeled with the number of the source given in Table
  \ref{obstable}.  The grayscale has
  a square-root transfer function spanning $f_\nu=0$ to 400 mJy
  beam$^{-1}$.}
\end{figure*}

The 193 targets were drawn, in order of precedence from (1) the
locations of millimeter cores in the Bolocam survey of the region
\citep{bolocam-perseus} (2) the locations of submillimeter cores in
the SCUBA survey of the region \citep{perseus-scuba}, (3) sources in
the literature survey of \citet{jijina}, and (4) cold,
high-column-density objects in the dust map produced by
\citet{schnee-mips}, and (5) weak detections that appear in both the
Bolocam and SCUBA maps but were not included in the published
catalogs.  Pairs of sources separated by less than than $31''$ (the
GBT beam FWHM) were reexamined and a single source was selected for
observation based on (sub)millimeter brightness.  Table
\ref{source-summary} summarizes the number of sources in each category
and their detection fractions.  Many of the same submillimeter cores
are identified in both the SCUBA and the Bolocam surveys of the
region.  When analysis of the higher-resolution SCUBA map revealed
sub-structure within a core identified as a single object in the
Bolocam map, we omitted the Bolocam source and observed the
substructure identified in the SCUBA catalog.  The locations of the
sources are shown in Figures \ref{findchart1} and \ref{findchart2}.
Nearly all of the (sub)millimeter sources are detected in ammonia
emission with the exception of NH3SRCs 143 and 184
\citep[BOLOCAM sources 90 and 120, ][]{bolocam-perseus}.  Both of
these millimeter sources are marginal detections.  We detect ammonia
at several positions without significant millimeter emission, notably
along 23 of the lines of sight selected based on their dust emission
in the far infrared.  Typical detections range from 0.5 to 4 K on the
$T_{mb}$ scale.  The noise levels in the spectra range from 40 to 150
mK, depending on integration time.  The noise values are determined
from the off-line regions of the spectra.  Off-line regions are
established iteratively as the regions more than 100 channels ($1.9
\mbox{ km s}^{-1}$) from significant (3$\sigma_{rms}$) emission.

\begin{deluxetable*}{cccccc}
\tablecaption{\label{source-summary} Summary of Source Origins}
\tablewidth{0pt}
\tablehead{
\colhead{Origin} & \colhead{Abbrev.} & \colhead{Number} &
\colhead{NH$_3$(1,1) Det. Frac.} & \colhead{NH$_3$(2,2) Det. Frac.} &
\colhead{C$_2$S Det. Frac.} 
}
\startdata
Bolocam & B & 115 & 98\% & 85\% & 67\% \\
SCUBA & S & 16 & 100\% & 94\% & 56\% \\
Literature & L & 5 & 100\% & 80\% & 100\% \\
Dust & D & 38 & 61\% & 8\% & 13\% \\
Weak Submm & W & 19 & 26\% & 11\% & 16\% \\
\hline
Total & \nodata & 193 & 84\% & 63\% & 51\% 
\enddata
\end{deluxetable*}

The C$_2 ^{34}$S line was not detected along any of the lines of
sight.  Given the ISM isotopic ratio $^{32}$S/$^{34}$S~$\sim 22$
\citep{ism-abund}, the lack of any detections,
particularly when the C$_2$S line is strong suggests that the main
line is usually optically thin.  For NH3SRC 42, we establish a lower
limit on the line ratio of 11.7 (the maximum in our population) by
setting the C$_2 ^{34}$S amplitude to the 3$\sigma$ limit.  Assuming
excitation conditions are the same for both isotopomers, this implies
$\tau_{\mathrm{CCS}}<1.4$ for one of the brightest C$_2$S lines in our
sample.

\begin{figure*}
\plotone{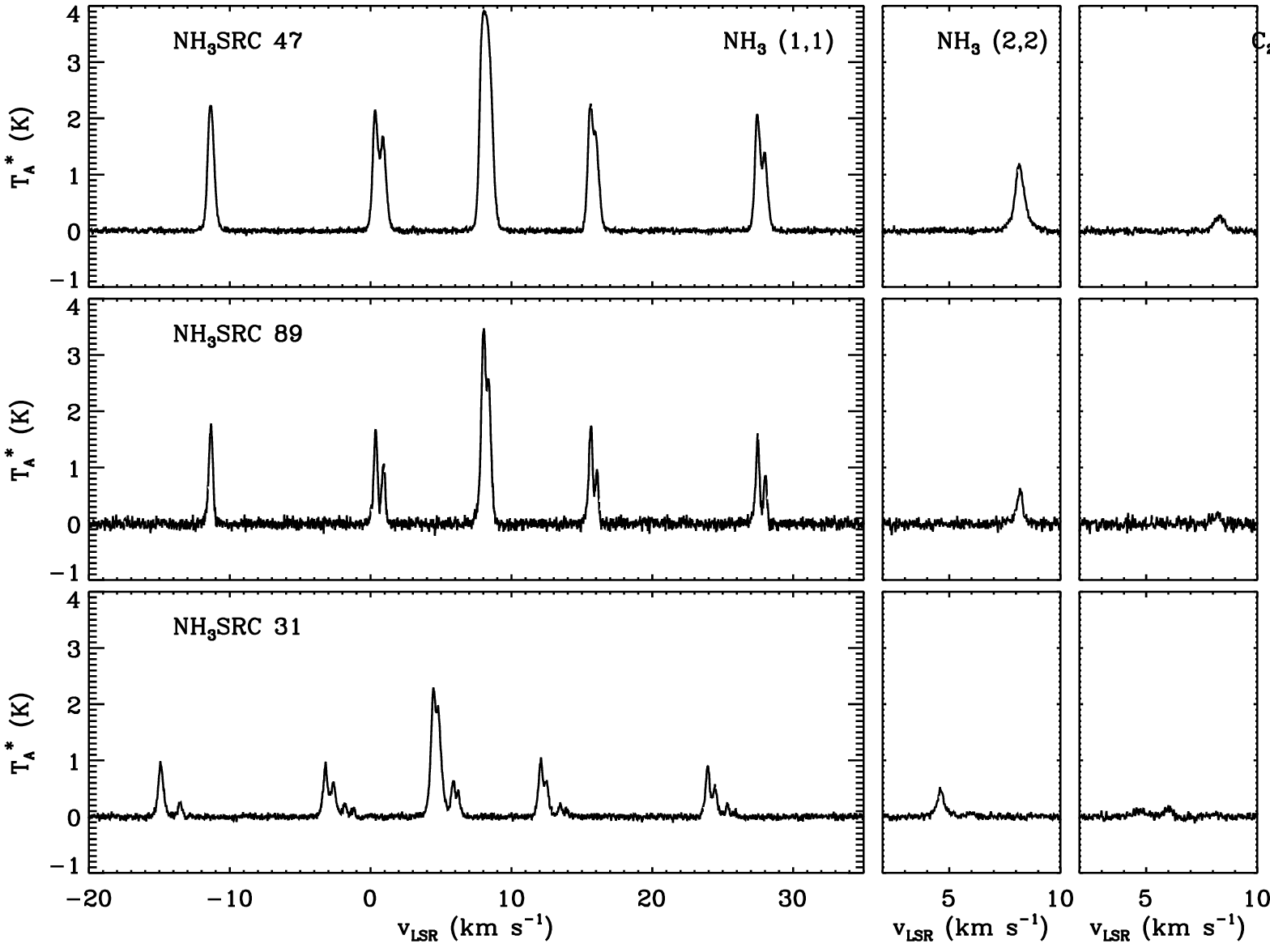}
\caption{\label{sampleplot} Three sample spectra from the ammonia
  sample including the source used for a flux check (Source 47), a
  narrow-line spectrum (Source 89) and a multicomponent spectrum
  (Source 31).}
\end{figure*}

In Figure \ref{sampleplot} we show three spectra from our sample.
Source 47 was observed every night as a consistency check on our flux
calibration and is thus the best observed ammonia source in our sample
(integrated S/N of 530).  Source 89 shows a typical narrow line
spectrum ($\sigma_v=0.12\mbox{ km s}^{-1}$) illustrating the
resolution of the data. Source 31 is a multi-component spectrum which
must be analyzed in more detail (see \S\ref{multicomp}). In Source 31,
the multiple components are also visible in the NH$_3$(2,2) and C$_2$S
lines.

We present a summary of our observations in Table \ref{obstable}.  The
properties of the cross-referenced names of the submillimeter cores
are given in \citet{bolocam-perseus} and \citet{perseus-scuba}
respectively.  Since ammonia has several hyperfine components that are
well-separated in velocity, the integrated intensities reported for
the (1,1) and the (2,2) lines are the sum of the integrated
intensities over all channels that are within 3~km~s$^{-1}$ of any
hyperfine component.  We subtract the mean intensity in the off-line
channels between the hyperfine components from the intensity of each
channel in the on-line windows to offset any low-lying baseline
residuals.  When there is no ammonia emission or C$_2$S from which the
line velocity can be determined, the main component of the ammonia
line is assumed to lie at the the mean $^{13}$CO velocity along the
line of sight, as derived from the COMPLETE $^{13}$CO data.

\section{Physical Parameter Estimation}
\label{params}
Here we describe the estimation of physical parameters from the
ammonia spectra.  The method differs somewhat from previous work in
that it forward models the properties of all observed spectral lines
simultaneously given input physical properties.  Then, the physical
properties are derived using a non-linear least squares minimization
code to determine the optimal fit to the observed spectrum.  This
stands in contrast with the standard method of obtaining these
properties which relies on measuring line ratios and using these line
ratios to calculate the physical properties.  While the results {\it
  should} be the same using the two methods, the primary advantage of
using a non-linear least squares fit is the automatic determination of
uncertainties in the derived physical parameters as well as the
covariances among those parameters (provided failures in the
assumptions of the least-squares problem are appropriately accounted for).
The model is optimized {\it simultaneously} for all observed spectra,
which eliminates many systematic effects that arise from comparing
properties derived from spectra separately.  An additional advantage
of this approach is that progressively more sophisticated models
\citep[e.g.][]{dvries05} can be introduced to model specific spectral
features.  However, for this survey, we adopt a relatively simple
model that can be applied to cores in a variety of environments.

For the spectra from a single object, a simple model is developed: the
emission is assumed to arise from a homogeneous slab with uniform gas
temperature, intrinsic velocity dispersion and uniform excitation
conditions for all hyperfine transitions of the NH$_3$ lines.
Detailed studies of NH$_3$ emission in conjunction with other
molecular tracers illustrate the shortcomings of this model.  The
observations of \citet{ladd-perseus} suggest that the ammonia emission
is not completely uniform on $30''$ scales.  \citet{tafalla04} note
that the kinetic temperature of the two cores they study in detail is
constant while there are radial variations in ammonia abundance and
excitation temperatures in cores.  \citet{mauersberger88} find the
line width of the NH$_3$(2,2) transitions to be larger than the
NH$_3$(1,1) transitions in the high-mass star forming region W3.

\begin{figure*}
\plotone{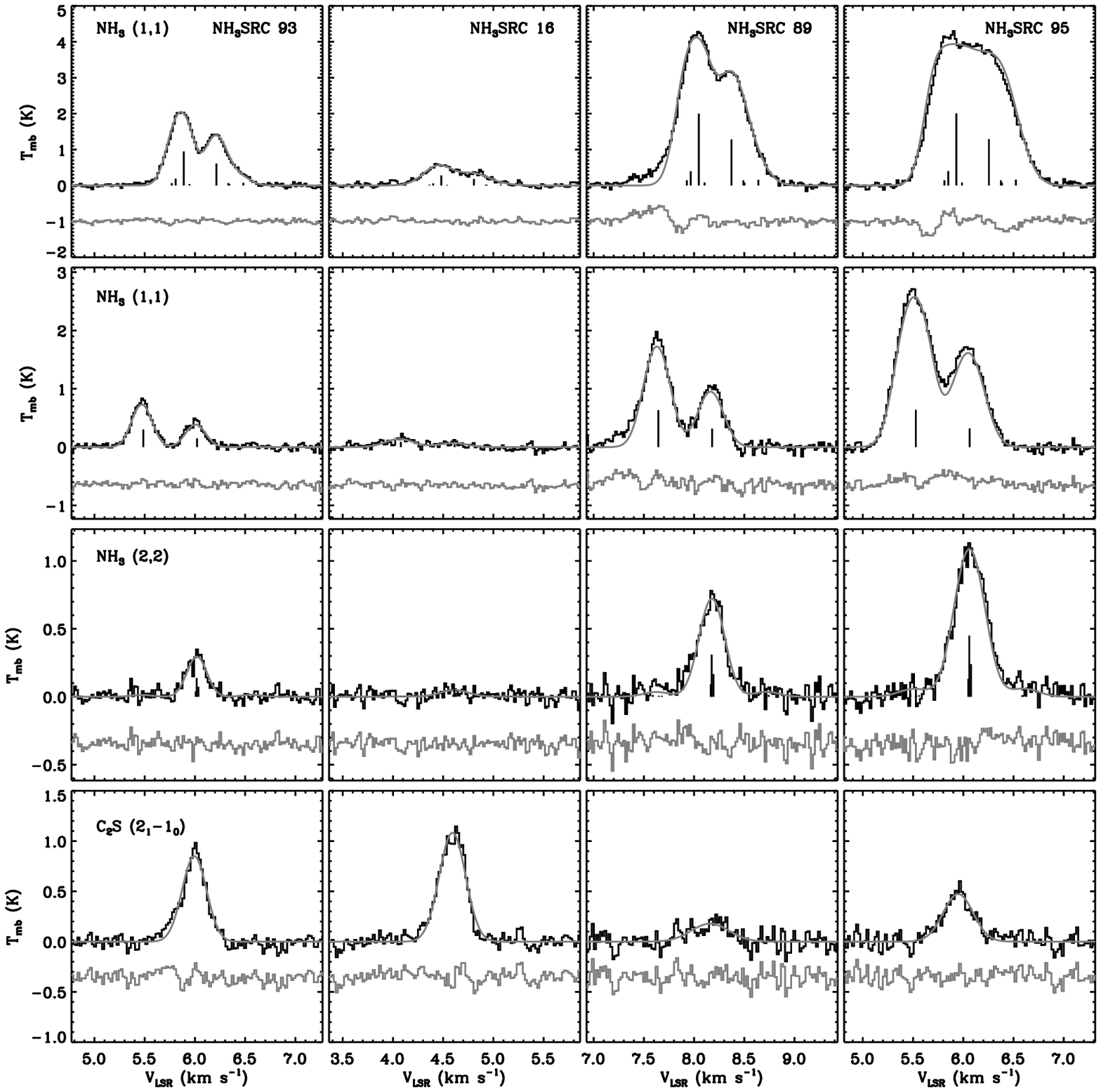}
\caption{\label{fitexample} Examples of fits to four GBT spectra.
  Each row shows a different excerpt from the spectra including (in
  order) the main component of the ammonia (1,1) complex, the
  satellite complex at $+19.85\mbox{ km s}^{-1}$ relative to the main
  complex (the $F_1,F=0,1/2\to 1,1/2$ transition; the $F_1,F=0,1/2\to
  1,3/2$ transition is also visible, offset by $-0.54$ km~s$^{-1}$),
  the main (2,2) complex and the C$_2$S line.  The gray, smooth line
  shows the model spectrum and the residual is shown below each
  profile.  Each column shows a different object including source 93,
  chosen for a good fit, source 16 chosen for a low optical depth fit,
  and sources 89 and 95 which show the slight deviations from the
  amplitude of the model in the case of high optical depth.  The
  velocities and relative strengths of the NH$_3$ hyperfine structure
  are indicated with vertical lines.}  
\end{figure*}

Despite these limitations, the uniform slab model is still useful.
Lacking information about the spatial distribution of ammonia, a
uniform slab is the simplest model we can adapt which should provide
reasonable average properties over the region within the beam
\citep[see, for example, the conclusions of][]{tafalla04}.  The GBT
beam is roughly half the typical core radius for objects in Perseus
\citet[\S\ref{observations}]{jijina}, so the emission should couple
well to the GBT beam (hence our adoption of the main beam temperature
scale).  Separate fits to the line width of the strong NH$_3$ (1,1)
and (2,2) detections show that the (2,2) line is, on average ($8\pm
2$)\% wider than the (1,1) line.  Several spectra show evidence for
non-uniform excitation of ammonia hyperfine components.  However, we
emphasize that these deviations in excitation and line width are small
(a few percent) and are only apparent because of the high quality of
the data.  In general, the slab model produces high-quality fits to
the data.  More complicated spectra (see Figure \ref{fitexample})
merit further investigation and these interesting sources will be
investigated in more detail elsewhere.  This presentation of the data
is restricted to generating average properties based on the simple
model for comparison of physical properties across the sample.

\subsection{Standard Spectral Model}

Our GBT reduction pipeline produces three spectra with 1.846 kHz
resolution centered on the NH$_3$ (1,1), NH$_3$ (2,2)
and C$_2$S ($2_1\to 1_0$) 
transitions.  The spectra are on the $T_A^*$ scale which include
corrections for atmospheric opacity.  We derive the physical parameters
for a simple ammonia system: the gas is assumed to have a slab
geometry with uniform properties, in particular gas kinetic
temperature.  The model assumes that column density of the material
has a Gaussian distribution in velocity with a dispersion of $\sigma_v$
around an LSR velocity centroid $v_{LSR}$:
\begin{equation}
\frac{dN}{dv} dv= N_0 \exp
\left[-\frac{(v-v_{LSR})^2}{2\sigma^2_v}\right]. 
\end{equation}
The ammonia (1,1) and (2,2) lines have 18 and 21 hyperfine components
respectively so the optical depth implied by the column density
distribution is split among each of the hyperfine components.  As
such, the opacity distribution for the (1,1) and (2,2) lines can be
written (in terms of frequency on the sky with respect to the LSR) as:
\begin{eqnarray}
\tau(\nu)& = &\tau_{1} \sum_{i=1}^{18} s_i
\exp\left[-\frac{(\nu-\nu_i-\nu_{LSR})^2}{2\sigma_i^2}\right]\nonumber
\\ &+&
\tau_{2} \sum_{j=1}^{21} s_j
\exp\left[-\frac{(\nu-\nu_j-\nu_{LSR})^2}{2\sigma_j^2}\right].
\label{fulltau}
\end{eqnarray}
where $\nu_{LSR}$ is determined by the Doppler formula (radio
convention):
\begin{equation}
\nu_{LSR}=\left(1-\frac{v_{LSR}}{c}\right)\nu_{rest}.
\end{equation}
Above, $\tau_1$ ($\tau_2$) is the total opacity in the (1,1) [(2,2)]
line transition; and $s_i$ ($s_j$) is statistical weight of the $i$th
($j$th) hyperfine component of the (1,1) [(2,2)] transition.  Each
component has a sky frequency $\nu_i$ ($\nu_j$) and a corresponding
width $\sigma_i$ ($\sigma_j$) given by
\begin{equation}
\label{doppler}
\sigma_i = \frac{\sigma_v}{c} \nu_i
\end{equation}
The number of molecules found in states that undergo the (1,1) vs.~the
(2,2) inversion transitions is governed by the {\it rotation}
temperature of the system ($T_R$).  Specifically, the population ratio
is established by the magnitude of $T_R$ relative to the energy gap
between the two states which (expressed in K) is $T_0=41.5$~K.  The
population ratio is given by the Boltzmann factor and the statistical
weights of the two states [$g{(1,1)}$=3 and $g{(2,2)}=5$].  We
assume that the transitions have equal line widths,
$\sigma_v(1,1)=\sigma_v(2,2)$, and excitation temperatures,
$T_x(1,1)=T_x(2,2)$.  After including the amplitudes of the dipole
matrix elements (e.g.~$|\mu(1,1)|^2$), the ratio of the opacities can
be expressed as \citep{ho79,nh3-araa}:
\begin{eqnarray}
\frac{\tau_2}{\tau_1}& =&
\left[\frac{\nu_{(2,2)}}{\nu_{(1,1)}}\right]^2
\frac{\sigma_v(1,1)}{\sigma_v(2,2)} \frac{T_x(1,1)}{T_x(2,2)}
\frac{|\mu(2,2)|^2}{|\mu(1,1)|^2}
\frac{g{(2,2)}}{g{(1,1)}}\exp\left(-\frac{T_0}{T_R}\right) \\ &=&
\left[\frac{\nu_{(2,2)}}{\nu_{(1,1)}}\right]^2
\frac{20}{9}\exp(-T_0/T_R).
\end{eqnarray}
We assume that the kinetic temperature $(T_k)$ is much less than $T_0$,
implying that the (1,1) and (2,2) states are the only populated
rotational levels of the ammonia system.  Thus, it is a two-state
system for which the kinetic temperature can be related to the
rotation temperature with knowledge of the collision coefficients
using detailed balance arguments \citep{swift-nh3}.  In this case:
\begin{equation}
T_R = T_k \left\{1+\frac{T_k}{T_0} \ln \left[1+
0.6\exp\left(-15.7/T_k\right)\right]\right\}^{-1}.
\end{equation}
Finally, given a radiation excitation temperature, $T_x$, the two
ammonia spectra can now be modeled in their entirety:
\begin{equation}
T_A^*(\nu) =  \eta_{mb}(\nu)\eta_f \left[J(T_x)-J(T_{bg})\right]
\left[1-e^{-\tau(\nu)}\right]
\label{spectrum}
\end{equation}
where $\eta_{mb}(\nu)$ is the main beam efficiency of the GBT
($\approx 0.81$ for the frequency range in this study), $\eta_f$ is the
filling fraction of the emission in the beam, $T_{bg}=2.73$~K and
\begin{equation}
J(T)=\frac{h\nu}{k} \frac{1}{\exp(h\nu/kT)-1}.
\end{equation}
After making assumptions about source-beam coupling and filling
fraction, the spectrum is entirely determined by five parameters:
$T_k, T_{x}, \tau_{1}, \sigma_v,$ and $v_{LSR}$.  Alternatively, we
can assume that $T_x=T_k$ (LTE) and let the filling fraction $\eta_f$
vary.

In addition to the ammonia system, we also measure spectra for the
C$_2$S line. We also fit a three parameter Gaussian to the C$_2$S line
simultaneously while deriving parameters for the ammonia spectra. The
total spectral model is given by:
\begin{equation}
T_A^* (\mbox{model}) = T_A^*(\mbox{NH}_3) + \eta_{mb}(\nu) T_{\mathrm{CCS}} \exp 
\left[-\frac{(\nu-\nu_0-\nu_{LSR}-\nu_{off})^2}
{2\sigma_{\mathrm{CCS}}^2}\right]
\end{equation}
Here, $\nu_0$ is the given rest frequency of the C$_2$S line,
$\nu_{LSR}$ is the appropriate frequency shift account for motion with
respect to the LSR using the same LSR velocity as derived from NH$_3$.
We define a velocity offset $V_{off}$ which, via the Doppler formula,
relates to the derived frequency offset
$\nu_{off}$. $T_{\mathrm{CCS}}$ and $\sigma_{\mathrm{CCS}}$ are the
amplitude and derived width of the line.  We do not know the
excitation temperature of the C$_2$S line, but we assume the excitation
is similar to that of ammonia since the critical density
\citep[$n_{cr}\sim 10^{4.5}\mbox{cm}^{-3}$,][]{langer95} is close to
that of the ammonia lines \citep{swade89}. The C$_2$S line complex
adds the parameters $T_{\mathrm{CCS}}$, $\sigma_{\mathrm{CCS}}$ and $
V_{off}$ to our fit.

The final effect we account for is the sampling of the GBT correlation
spectrometer.  The lags in the spectrometer are weighted uniformly
(i.e.~no online smoothing).  As a result, a channel has a nominal
profile of a sinc function:
\begin{equation}
\phi(\nu) = \frac{\sin\left[\pi
    \left(\nu-\nu_k\right)/\Delta\nu\right]}{\pi
    \left(\nu-\nu_k\right)/\Delta \nu}.
\end{equation}
Here, $\nu_k$ is the channel center and $\Delta \nu$ is the channel
spacing.  After deriving $T_A^* (\mbox{model})$, we digitally sample
it at one-fourth the channel spacing and we convolve the model
spectrum with channel profile to produce the final result which is
compared with observations.

We find the maximum likelihood model for the spectrum using a
non-linear least-squares fitting routine\footnote{C. Markwardt's MPFIT
  package.  Note that, since the channels are not independent, strict
  least squares fitting is technically inappropriate.  See
  \S\ref{errors}}.  The optimization occurs in two steps: first the
fit is performed to the entirety of the NH$_3$(1,1), NH$_3$(2,2) and
C$_2$S spectra.  If emission is detected the fit is performed again,
ignoring regions of the spectrum more than 2.5~km~s$^{-1}$ away from
significant emission using the results of the first fit as an initial
guess for the optimization.  This second step reduces the number of
noise-only channels in the fit and allows for a cleaner convergence to
an optimal set of properties.  Examples of the fitting appear in
Figure \ref{fitexample}.  We have chosen four representative spectra
for several cases found in the single-component models.  The first two
columns of the figure show the successful applications of the model in
the regular and low optical depth regimes (\S\ref{locol}).  The final
two columns of the figure show the slight deviations frequently
encountered in spectra with high optical depth in the lines.  The
slight deviations may be the result of non-LTE excitation of the
different hyperfine components.  Several other spectra show
asymmetries in the line profiles for which a Gaussian model of the
velocity distribution is inaccurate (see \S\ref{multicomp} for further
discussion of these cases).  The full set of spectra for lines of
sight with detections are available as online-only figures (Figure
\ref{fullspex}--6ff) and are downloadable from the COMPLETE
website\footnote{\url{http://www.cfa.harvard.edu/COMPLETE/data\_html\_pages/GBT\_NH3.html}}.

\subsection{Low Optical Depth Regime}
\label{locol}
When $\tau(\nu)\ll 1$ for the ammonia complex over all $\nu$, the
parameters $T_{x}$ and $\tau_{11}$ become degenerate and it is
impossible to solve for the two parameters independently.  In this
case, we expand Equation \ref{spectrum} assuming the Rayleigh-Jeans
limit so that 
\begin{equation}
T_A^*(\nu) =  \eta_{mb}(\nu)\eta_f \left(T_x- T_{bg}\right) \tau(\nu)
\label{lowopt}
\end{equation}
where $\tau(\nu)$ is given by Equation \ref{fulltau}.  In this case,
we optimize the fit for the free parameter
$\gamma\equiv\left(T_x-T_{bg}\right)\tau_1$ and the ammonia spectrum
is determined by four free parameters: $T_k, \gamma, \sigma_v$, and
$v_{LSR}$.  This approximation is accurate to better than 10\% for
all components of the ammonia complex provided $\tau_1<1$.

\subsection{Column Density Estimates}
We measure the total column density of NH$_3$ and C$_2$S using the
derived parameters from the fit.  For example, the column density in
the NH$_3$ (1,1) state is \citep[{\it e.g.}\rm][]{rohlfs-wilson}: 
\begin{eqnarray}
N(1,1) =\frac{8\pi\nu_0^2}{c^2}\frac{g_1}{g_2}\frac{1}{A_{(1,1)}}
\left[1-\exp \left(\frac{h\nu_0}{kT_{x}}\right)\right]^{-1} \int \tau(\nu) d\nu
\label{columndens}
\end{eqnarray}
The statistical weights of the upper and lower levels of the inversion
transition are equal.  The Einstein A value for the inversion
transition is $A_{(1,1)}=1.68\times 10^{-7}\mbox{ s}^{-1}$
\citep{jplcat}.  The opacity per unit frequency is given in terms of
the total opacity of the line ($\tau_1$) by the first term in Equation
\ref{fulltau}.
\begin{eqnarray}
 \int \tau(\nu) d\nu & = &\tau_1 \int d\nu \sum_i s_i 
\exp\left[-\frac{(\nu-\nu_i-\nu_{LSR})^2}{2\sigma_i^2}\right]\nonumber \\
& = &
\sqrt{2\pi}\sigma_i\tau_1 = \sqrt{2\pi}\frac{\sigma_v}{c}\nu_0 \tau_1
\end{eqnarray}
under the approximation that the frequencies of the individual
hyperfine components are the same ($\nu_0$). The partition function
for the metastable states ($J=K$) of ammonia is given by
\citep{rohlfs-wilson}:
\begin{equation}
Z =\sum_J (2J+1)~S(J)~\exp\left\{\frac{-h[B~J(J+1)-(C-B)J^2]}{k T_{K}}\right\}
\end{equation} 
Here, $B$ and $C$ are the rotational constants of the ammonia
molecule: 298117~MHz and 186726~MHz respectively \citep{jplcat}.  The
factor $S(J)$ equals 2 for $J=3,6,9\dots$ and 1 otherwise, accounting
for the extra statistical weight of ortho-NH$_3$ over
para-NH$_3$. Owing to the relatively short lifetime of the $J\neq K$
states, we assume all ammonia molecules are in the metastable states.
To determine the total column density, we scale the column density in
the (1,1) state by $Z/Z(J=1)$.  We truncate the partition function at
50 terms.

For the low optical depth case, we expand the exponential in Equation
\ref{columndens}:
\begin{eqnarray}
N(1,1)& = &\frac{8\pi k \nu_0}{h c^2}\frac{1}{A_{(1,1)}}
\int T_x \tau(\nu) d\nu \nonumber\\
& = &\frac{8\pi k \nu_0^2}{h
  c^3}\frac{1}{A_{(1,1)}} \sqrt{2\pi} \sigma_v~  (T_x-T_{bg}) \tau_1
\label{lowop}
\end{eqnarray}
where $\sigma_v$ and   $(T_x-T_{bg})\tau_1$ are determined by the
low-optical-depth fit (\S\ref{locol}).  

We also estimate the column density of C$_2$S assuming that the line
is optically thin, an assumption bolstered by our failure to detect
C$_2$$^{34}$S along any of the lines-of-sight.  
\begin{equation}
N(\mathrm{C_2S},J=1) = \frac{8\pi k \nu_0}{h c^2}\frac{g_1}{g_2}\frac{1}{A_{21}}
 \sqrt{2\pi} \sigma_v~  (T_x-T_{bg}) \tau_{21}.
\label{CCS}
\end{equation}
Here, $g_J=2J+1$, $A_{21}=5.44\times 10^{-7}\mbox{ s}^{-1}$
\citep{jplcat} and we take $(T_x-T_{bg}) \tau_{21}=T_{\mathrm{CCS}}$.  Again, we
calculate the total column density of C$_2$S using the partition
function.  For a
state with energy above the ground state $E_i$ and degeneracy $g_i$,
the partition function is the standard
\begin{equation}
Z = \sum_i g_i \exp\left[-\frac{E_i}{k T_k}\right].
\end{equation}
We adopt the values of $E_i$ and $g_i$ from the tabulated molecular
data in the JPL molecular spectroscopy catalog \citep{jplcat} and
assume the 295 states are thermally populated.  It may not be appropriate
to use the kinetic temperature derived from the NH$_3$ for the C$_2$S
partition function since the two species may not be thermally coupled.
This systematic effect limits our ability to measure the C$_2$S
column.  For both molecules, the uncertainties in the column density
are established by adding a normal deviate times the uncertainty to
the input parameters and recalculating the column densities.  After
repeating this redistribution within the errors a large number of
times, the error are determined from the width of the resulting
distribution.

We tested the results of the uniform slab modeling by comparing the
results to the values derived from the hyperfine fitting routines in
the CLASS package.  We checked the line width, opacity, excitation
temperature and LSR velocity.  The results were identical within the
errors of our analysis except for complex source spectra ({\it
e.g.}~asymmetric profiles, multiple components).

\subsection{Multi-Component Fitting}
\label{multicomp}
Several of the spectra show significant velocity structure in the line
beyond what is expected from the hyperfine structure of ammonia (see,
for example, Source 31 in Figure \ref{sampleplot}).  In cases where
the number of components is readily modeled, we have attempted to fit
a multicomponent model to the spectrum.  In this model, we assume that
there are two objects in the beam each with $\eta_f<1$ and we operate
in the LTE approximation ($T_x = T_k$).  Hence, it is not necessary to
calculate radiative transfer effects of one component through another.
The approximation appears to be sufficiently good for our purposes.

We only present multiple components to the data where the evidence is
unambiguous that a multiple-component fit is appropriate. This means
two clear peaks in the NH$_3$ (2,2) line. In some cases the velocity
separation is sufficient that multiple components are also well
distinguished in the (1,1) line. The initial conditions for the fit
are established by hand, but the optimization is performed
simultaneously for both components.  To prevent run-away solutions, we
constrain the initial velocities to be within 0.1 km~s$^{-1}$ of the initial
guess and also constrain the C$_2$S and NH$_3$ velocities to be within
0.1 km~s$^{-1}$ for each of the components. We report the fits to the
multiple components independently in Table \ref{proptable} appending a
decimal and the number of the component onto source name.  In the six
spectra with multiple components reported, all show that the sum of
the their filling fractions is less than unity.  The typical filling
fraction for a single component is $\eta_f=0.3$.

Seven additional spectra show strong evidence for multiple components
in some or all of the lines. In particular, there are often multiple
components in the C$_2$S and a broad or poorly fit NH$_3$ (1,1) line
but no clear evidence for multiple components in the NH$_3$ (2,2)
line. Since we cannot be certain that these two components in C$_2$S
are physically associated with two components in NH$_3$, we refrain
from performing a multicomponent fit but note the presence of the
components in Table \ref{proptable}. There are a number of additional
spectra with weaker evidence of multiple components, or evidence for
more than two components. We likewise flag these objects in Table
\ref{proptable}.

\subsection{Uncertainties and Limits in Derived Parameters} 
\label{errors}

The reported uncertainties in Table \ref{proptable} are the derived
uncertainties from the nonlinear least-squares fitting using the
covariance matrix.  The individual channels are not independent (the
channel width is 1.525 kHz and the resolution is 1.862 kHz) which
violates an assumption of least-squares fitting.  The errors
determined reported from the covariance matrix represent the $\Delta
\chi^2=1$ ellipsoid.  To determine appropriate uncertainties, we
generated multiple realizations of several different spectral models
each with different noise distributions.  We used our fitting routine
to derive parameters in the presence of noise and compared the
distribution of the derived parameters to the input model parameters.
Over all the different spectral models, the true errors were a factor
of $\le 1.6$ larger than the errors derived from the covariance matrix
assuming the data were independent.  While not technically a
least-squares fit, accurate confidence intervals for the derived
parameters can still be estimated using the $\Delta \chi^2=2.56=1.6^2$
surface \citep{numrec}.  We have investigated the shape of $\chi^2$
space in our data and find that the region around the minimum $\chi^2$
is well approximated by a paraboloid.  To report errors consistent
with a $1\sigma$ spread around the derived parameters, we scale the
reported errors up by a factor of 1.6.

The covariance matrix also indicates which parameters are correlated
with each other in the fits.  The additional uncertainty due to this
correlation is accounted for in the reported errors.  For parameters
$i$ and $j$, we express the covariance in terms of the normalized
covariance: $\sigma_{ij}/(\sigma_i \sigma_j)^{1/2}$.  We examined the
average covariance matrix over the 133 fits with sufficiently strong
NH$_3$(1,1) emission to fit for optical depth and excitation
temperature separately.  We find that the most obvious
(anti)correlation is between $T_{ex}$ and $\tau$:
$\sigma_{ij}/(\sigma_i \sigma_j)^{1/2}=-0.95$.  This strong
anticorrelation necessitates the low opacity treatment
described in \S\ref{locol}.  The excitation temperature is also
correlated with $T_{K}$ (0.18) and $\sigma_v$ (0.24).  The line
opacity ($\tau_i$) has an anticorrelation with the line width
$\sigma_v$ ($-0.49$) and $T_k$ ($-0.23$).  For the C$_2$S line, there is
the anticorrelation between the amplitude ($T_{\mathrm{CCS}}$) and the
line width $\sigma_{\mathrm{CCS}}$ typical of Gaussian fits ($-0.57$ in
our case). The other elements of the covariance matrix are consistent
with zero.

%% \begin{deluxetable}{llllllll}
%% \tablecaption{\label{cvarr} Normalized Covariance Array}
%% \tablewidth{0pt}
%% \tablehead{\colhead{$T_{kin}$} & \colhead{$\tau_{1}$} & \colhead{$\sigma_v$} &
%% \colhead{$v_{lsr}$} & \colhead{$T_{ex}$} &
%% \colhead{$T_{\mathrm{CCS}}$} & \colhead{$v_{off}$} & \colhead{$\sigma_{\mathrm{CCS}}$}
%% }
%% \startdata
%% 1.00 & -0.23 & 0.11 & 0.01 & 0.18 & \nodata & \nodata & \nodata\\
%% -0.23 & 1.00 & -0.49 & -0.04 & -0.95 & \nodata & \nodata & \nodata\\
%% 0.11 & -0.49 & 1.00 & 0.10 & 0.24 & \nodata & -0.01 & \nodata\\
%% 0.01 & -0.04 & 0.10 & 1.00 & 0.02 & \nodata & -0.08 & \nodata\\
%% 0.18 & -0.95 & 0.24 & 0.02 & 1.00 & \nodata & \nodata & \nodata\\
%% \nodata & \nodata & \nodata & \nodata & \nodata & 1.00 & \nodata & -0.57\\
%% \nodata & \nodata & -0.01 & -0.08 & \nodata & \nodata & 1.00 & \nodata\\
%% \nodata & \nodata & \nodata & \nodata & \nodata & -0.57 & \nodata & 1.00\\
%% \enddata
%% \end{deluxetable}

The uncertainties do {\it not} reflect the overall
uncertainty in the amplitude scale calibration which is $\sim 5\%$,
The relative calibration across the spectra is much better than $5\%$,
so certain properties are unaffected by the overall amplitude
calibration.  The integrated intensities reported in Table
\ref{obstable} as well as properties derived from the amplitude of the
emission (column densities, excitation temperatures, filling
fractions, and antenna temperatures) have $\sim 5\%$ uncertainties.
In contrast, line-of-sight velocities, line widths, optical depths and
kinetic temperatures (since the latter two are driven by line ratios)
have uncertainties close to their reported precisions.  

In several cases, we detect the (1,1) transition of NH$_3$ but not the
(2,2) transition.  When we can establish an upper limit on the
intensity of (2,2) line, we report a $3\sigma$ upper limit on the
temperature of the ammonia.  Since the ammonia temperature is used in
the calculation of column densities, the upper limit on temperature
produces a {\it lower} limit on the NH$_3$ column density since the
partition function correction is a decreasing function of temperature
for $T_k<40$~K.  In contrast, the correction for C$_2$S is an
increasing function of temperature so the upper limit of temperature
creates an upper limit for the C$_2$S column density.  

The velocity width of the ammonia complex is quoted as an upper limit
in Table \ref{proptable} for instances when (1) there are multiple
components along the line of sight that cannot be decoupled and fit
separately or (2) when the line widths are large but the
signal-to-noise is small such that the broadening of the line cannot
be distinguished from the splitting due to the hyperfine structure.
In the latter case, the upper limit reported may represent the actual
value, but we cannot distinguish between a large intrinsic line width
and line widths that result from the hyperfine structure.  We note
that we find no cores with large ammonia (1,1) antenna temperatures
($T_A^* > 1\mbox{ K}$), large line widths ($\sigma_v > 0.2\mbox{ km
  s}^{-1}$) which have no evidence for multiple components along the
line of sight.  Said differently, all large line width cores {\it may}
have large line widths only because of multiple components along the line
of sight.

The largest systematic in the reported values is the bias introduced
by the uniform slab model presented above.  Again, we emphasize that
the model is adopted for uniform application to a large sample;
individual spectra can be investigated in more detail.  One difficulty
in applying our model may occur in comparing the (1,1) and (2,2)
emission.  Although the critical densities for the two transitions are
similar, the NH$_3$(1,1) emission has larger optical depths than the
(2,2) emission.  In cores with radial gradients in temperature,
opacity effects may result in the (2,2) emission revealing warmer gas
than the (1,1) emission.  We have attempted fitting the most optically
thick spectra with the hill models of \citet{dvries05}, but we do not
find a significant improvement in the fit quality with the additional
complications the model entails.  Such line-of-sight variations in the
excitation conditions are ignored in our simple treatment, but our
derived values should yield (appropriately weighted) average
conditions along the line of sight.

\subsection{Comparison to Previous Work}

Ammonia has been observed towards Perseus in several previous studies.
We compare the derived properties from our analysis to those values
found in the literature for sources other studies have observed.  For
comparison, we use homogenized properties in the catalog of
\citet{jijina} which are drawn primarily from the work of
\citet{ladd-perseus,bachiller86,bachiller87} and \citet{juan93}.  We
find good agreement between our line widths and temperatures with
$\lesssim 20\%$ variations across most of our sources. Discrepant
points are invariably found in NGC 1333 where larger line widths are
typically found in earlier studies.  We suspect that the larger beam
sizes of previous work blend together more disparate emission than the GBT
observations resulting in the larger line widths.  The agreement with
ammonia column density is less well established with variations up to
0.5 dex are found.  However, the largest discrepancies are associated
with highly uncertain column densities flagged in \citet{jijina}.  We
conclude that these new data match the results of previous studies
quite well with significant variations attributable to the improved
quality of the observations.

\section{Distributions of Derived Properties}
\begin{figure*}
\plotone{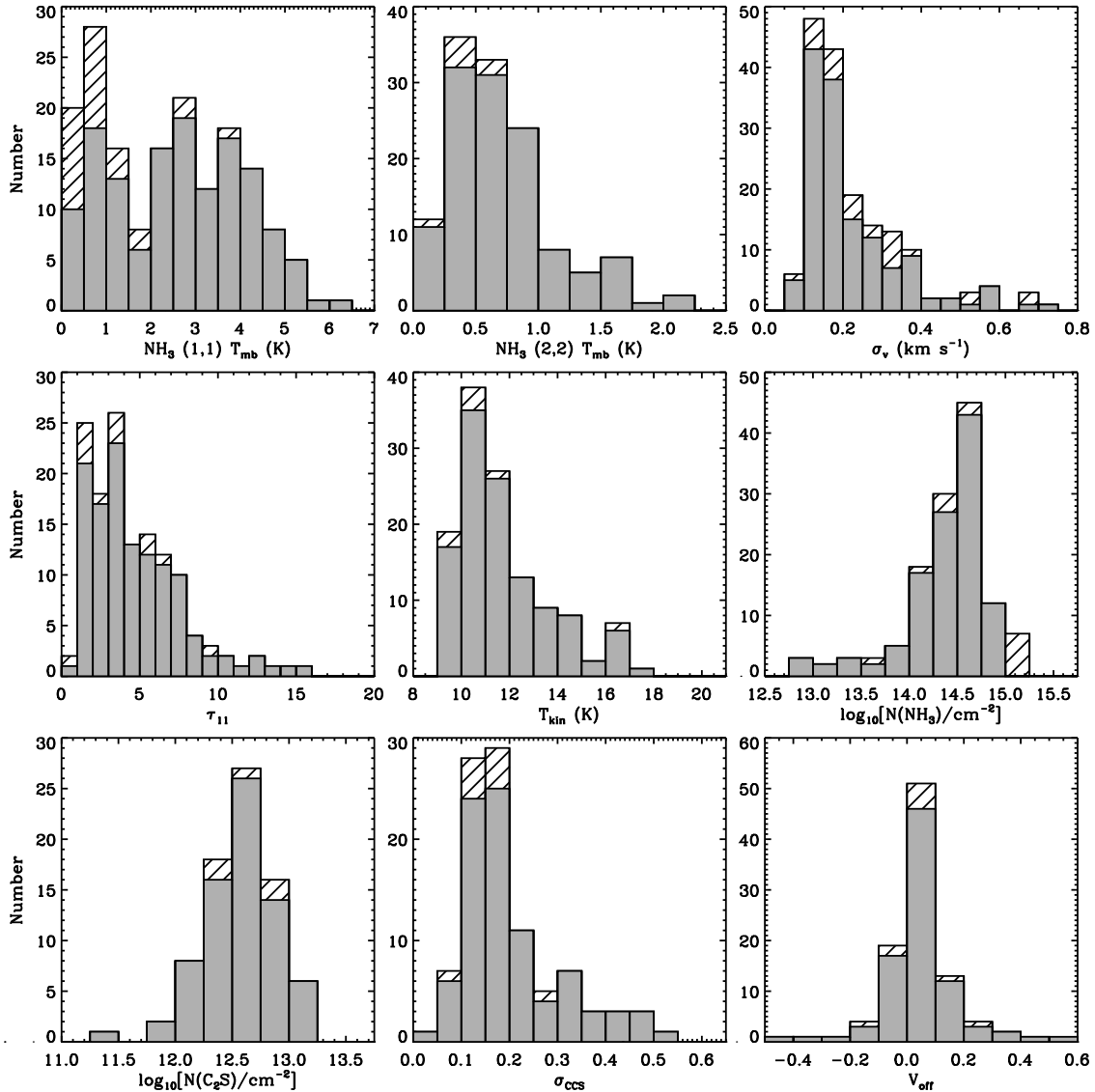}
\caption{\label{propdist} Distribution of the observed and derived
  properties for the GBT spectra.  Working from left to right, top to
  bottom, the nine panels show the peak main beam temperatures for the
  (1,1) and (2,2) lines, the intrinsic velocity dispersion of NH$_3$
  derived from the (1,1) and (2,2) lines (including upper limits as
  data), the total optical depth of the (1,1) complex, the kinetic
  temperature of the gas, the column density of the ammonia emission,
  the column density of the C$_2$S, the velocity dispersion of the
  C$_2$S line, and the velocity offset of the C$_2$S line from the
  ammonia complex.  The gray histogram shows the distributions of
  properties for the pointings associated with (sub)millimeter sources
  and the hashed histogram shows all remaining objects.  Objects are
  included in the histograms only if they have well-determined values
  for the properties reported (with the exception of line width).}
\end{figure*}

In this section, we present a brief summary of the observed spectra
and their derived properties.  A graphical summary of the data that
appear in Tables \ref{obstable} and \ref{proptable} is given in Figure
\ref{propdist}.  The first two panels show the typical distribution of
line temperatures on the $T_{mb}$ scale for the (sub)millimeter (gray)
and all other sources.  The strongest sources are all associated with
millimeter-bright objects and other targets are typically weak in
(1,1) emission and infrequently detected in (2,2).  Two sources,
NH3SRC 27 and 60, are outside the bounds of either (sub)millimeter
study but we detect significant line emission.  We have included these
in the millimeter-faint population since they are only associated
with MIPS-derived dust features, but this assignment may be incorrect.

The derived intrinsic line widths are typically $<0.2$~km s$^{-1}$
across the entire sample with a high line width tail to the
distribution.  As noted previously, these line widths may be upper
limits since in all cases where there is sufficient signal-to-noise to
resolve the structure of the line there is evidence for multiple
velocity components.  The typical (total) optical depth of the ammonia
complex is $\sim 4$ and the main complex has a thickness half the
optical depth shown.  Hence, in most cases, the lines are only
moderately opaque, though some line complexes are quite optically
thick.  The (sub)millimeter-weak sources have a higher median line
width and lower opacity than the (sub)millimeter-bright population.

The derived kinetic temperatures of the cores are uniformly cool
(T$_K<20$~K) and are typically 11~K, substantially lower than is
assumed in some work \citep[e.g.][]{perseus-scuba} for submillimeter
cores.  If the dust and gas are well-coupled, assuming $T=15$~K for
the Perseus cores can result in underestimating the mass of the cores
by a factor of 1.7.  However, assuming a temperature of 10~K yields a
typical overestimate by a factor of 1.2.  To accurately determine the
masses of cores from the millimeter continuum requires temperature
determinations for every core.  After this correction, the dominant
contribution to the uncertainties in the core masses is the dust
opacity at these wavelengths.

The column density of ammonia is typical for cores in Perseus
\citep{jijina}.  However, the sensitive observations also find some
spectra that imply $N(\mbox{NH}_3)< 10^{13.5}\mbox{ cm}^{-2}$, making
these detections among the lowest column densities of ammonia yet
found.  The low column density detections are all associated with the
IC348 region of the cloud.  The C$_2$S column densities also appear
typical of dense cores \citep{suzuki92} but are subject to uncertainty
based on unknown temperatures and excitation conditions.

The velocity dispersions of the C$_2$S lines are comparable to those
of the ammonia lines, and the C$_2$S lines show a slight, systematic
offset in velocity from the ammonia complex.  This offset is likely
due to uncertainties in the assumed rest frequency of the C$_2$S line.
The mean offset is 16 m~s$^{-1}$ (weighting by the inverse variance of
the measurements) and would be consistent with zero for a rest
frequency of $\nu_{\mathrm{CCS}}=22.344032(1)$~GHz.  The difference is
within the uncertainties of the assumed frequency.

We conclude this section by noting several spectra that define the
extent of the property distributions or are otherwise notable.  Plots
of the spectra are available in the online-only edition (Figure
\ref{fullspex}--6ff).

{\it Typical Spectrum} -- NH3SRC 15 is the ``most typical'' ammonia
spectrum from Perseus with nearly average values of all the properties
shown in Figure \ref{propdist}.  For NH3SRC 15, $T_{kin}=11.2$~K,
$\sigma_v=0.19$~km~s$^{-1}$, $\tau_{1,1}=3.7$,  and
$N(\mbox{NH}_3)=3.1\times 10^{14}\mbox{ cm}^{-2}$.

{\it Temperature Range} -- NH3SRC 18 has the lowest,
well-determined temperature of the observed sources ($T_{kin}=9.05$~K)
and NH3SRC 116 has the highest temperature ($T_{kin}=26$~K).

{\it Column Density} -- NH3SRC 144 has the lowest column density
detected in our survey ($5.5\times 10^{12}$~cm$^{-2}$) and NH3SRC 17
has the highest column density ($1.3\times 10^{15}$~cm$^{-2}$). 

{\it Line Brightness} -- NH3SRC 54 is the faintest source in
NH$_3$(1,1) emission included as a detection
($W_{mb}=0.2$~K~km~s$^{-1}$) and NH3SRC 12 is the strongest
(20.0~K~km~s$^{-1}$).  In the (2,2) line, NH3SRC 19 is the weakest
($W_{mb}=0.03$~K~km~s$^{-1}$) and NH3SRC 68 is the strongest
($W_{mb}=2.27$~K~km~s$^{-1}$).  In the C$_2$S line, NH3SRC 21 is the
weakest detection ($W_{mb}=0.07$~K~km~s$^{-1}$) while NH3SRC 42 is the
strongest ($W_{mb}=0.67$~K~km~s$^{-1}$).

{\it Line Width} -- The narrowest line width source we detect is
NH3SRC 128 with a line width of 0.079~km~s$^{-1}$.  The largest line
width we reliably detect is 0.23~km~s$^{-1}$ in NH3SRC 109.  However,
many of the fits yield larger results such as NH3SRC 71 where the
measured line width is 0.72~km~s$^{-1}$ though the fit is unreliable.
In addition many spectra show odd structure in their line profiles
including wings (NH3SRCs 70, 127) and plateaus (NH3SRC 75) in addition
to the multicomponent structure discussed previously
(\S\ref{multicomp}).

\section{Summary}  
We have searched for NH$_3$(1,1), NH$_3$(1,1), C$_2$S($2_1\to 1_0$)
emission along 193 lines of sight towards the Perseus molecular
cloud.  The lines of sight were selected based on positions that were
detected in (sub)millimeter emission or had large dust column densities
implied by far infrared (FIR) emission.  We detect ammonia emission along
162 (84\%) of the lines of sight and C$_2$S along 96 (51\%) of the
lines of sight.  We estimate the physical properties of the gas by
fitting a model emission profile to all spectral lines simultaneously.
The emission is modeled as a uniform slab of gas that completely fills
the beam, has a Gaussian intrinsic line width, and a single
excitation temperature for all lines.  Where appropriate, we refined
the model to account for low optical depths, incomplete coupling to
the GBT beam and multiple velocity components along the line of sight.

Nearly all (98\%) bright, (sub)millimeter cores have strong ammonia
emission associated with them and the exceptions appear to be
artifacts in the submillimeter map based on examining the original
BOLOCAM data.  In addition, we detected emission towards 23 sources
selected based on FIR emission that implies large dust column
densities and low temperatures.  Twenty-one objects are not seen in
the (sub)millimeter, suggesting that the submillimeter emission is not
a perfect tracer of the dense gas (the remaining two sources are
outside the bounds of the continuum surveys).  However, the FIR-based
ammonia detections have lower line intensities than
(sub)millimeter-bright source, as well as lower optical depths and
larger line widths.  It remains to be shown whether this could be an
evolutionary effect or whether the (sub)millimeter-weak sources simply
trace isolated pockets of gas not associated with the dense cores
traced by the dust continuum.

We find that the ammonia implies dense gas temperatures in Perseus are
predominantly cold ($T_k\sim 11~$K).  Ammonia column densities are
typical for cores presented in the literature
\citep[$N_{\mathrm{NH3}}\sim 10^{14.5}\mbox{ cm}^{-2}$,][]{jijina}
though we also find several lines-of-sight with very low ammonia
column densities ($N_{\mathrm{NH3}}\lesssim 10^{13.5}\mbox{ cm}^{-2}$)
associated with the IC 348 region.

Forthcoming work will examine the properties of these objects in more
detail including comparison with the (sub)millimeter emission,
protostellar content, and the velocity structure of the dense core
population.

\acknowledgements The Green Bank Telescope is operated by the National
Radio Astronomy Observatory.  The National Radio Astronomy Observatory
is a facility of the National Science Foundation operated under
cooperative agreement by Associated Universities, Inc.  We are
grateful for the assistance of Ron Maddalena and Frank Ghigo in
preparation for the observations and to the GBT operators who executed
the observations.  We acknowledge the indispensable assistance and
advice of Scott Schnee, Doug Johnstone, Melissa Enoch and Helen Kirk
in the planning of and preparation for the observations, particularly
for the use of their continuum maps.  ER's work is supported by an NSF
Astronomy and Astrophysics Postdoctoral Fellowship (AST-0502605).  JEP
and JBF are supported by a generous grant from the NRAO Student
Observing Support Program (GSSP06-0015).  JEP is supported by the
National Science Foundation through grant \#AF002 from the Association
of Universities for Research in Astronomy, Inc., under NSF cooperative
agreement AST-9613615 and by Fundaci\'on Andes under project
No. C-13442. This material is based upon work supported by the
National Science Foundation under Grant No. AST-0407172.  PC
acknowledges supported by the Italian Ministry of Reserach and
University within a PRIN project.

{\it Facilities:} \facility{GBT (K-band/ACS)}

%% \bibliographystyle{apj} 
%% \bibliography{refs}

\begin{thebibliography}{35}
\expandafter\ifx\csname natexlab\endcsname\relax\def\natexlab#1{#1}\fi

\bibitem[{{Bachiller} \& {Cernicharo}(1986)}]{bachiller86}
{Bachiller}, R. \& {Cernicharo}, J. 1986, \aap, 168, 262

\bibitem[{{Bachiller} {et~al.}(1987){Bachiller}, {Guilloteau}, \&
  {Kahane}}]{bachiller87}
{Bachiller}, R., {Guilloteau}, S., \& {Kahane}, C. 1987, \aap, 173, 324

\bibitem[{{Cernis}(1993)}]{cernis93}
{Cernis}, K. 1993, Baltic Astronomy, 2, 214

\bibitem[{{De Vries} \& {Myers}(2005)}]{dvries05}
{De Vries}, C.~H. \& {Myers}, P.~C. 2005, \apj, 620, 800

\bibitem[{{Di Francesco} {et~al.}(2006){Di Francesco}, {Evans}, {Caselli},
  {Myers}, {Shirley}, {Aikawa}, \& {Tafalla}}]{pp5-difran}
{Di Francesco}, J., {Evans}, N.~J., {Caselli}, P., {Myers}, P.~C., {Shirley},
  Y., {Aikawa}, A., \& {Tafalla}, M. 2006, ArXiv Astrophysics e-prints

\bibitem[{{Enoch} {et~al.}(2006){Enoch}, {Young}, {Glenn}, {Evans}, {Golwala},
  {Sargent}, {Harvey}, {Aguirre}, {Goldin}, {Haig}, {Huard}, {Lange},
  {Laurent}, {Maloney}, {Mauskopf}, {Rossinot}, \& {Sayers}}]{bolocam-perseus}
{Enoch}, M.~L., {Young}, K.~E., {Glenn}, J., {Evans}, N.~J., {Golwala}, S.,
  {Sargent}, A.~I., {Harvey}, P., {Aguirre}, J., {Goldin}, A., {Haig}, D.,
  {Huard}, T.~L., {Lange}, A., {Laurent}, G., {Maloney}, P., {Mauskopf}, P.,
  {Rossinot}, P., \& {Sayers}, J. 2006, \apj, 638, 293

\bibitem[{{Flower} {et~al.}(2006){Flower}, {Pineau Des For{\^e}ts}, \&
  {Walmsley}}]{flower06}
{Flower}, D.~R., {Pineau Des For{\^e}ts}, G., \& {Walmsley}, C.~M. 2006, \aap,
  456, 215

\bibitem[{{Hatchell} {et~al.}(2007){Hatchell}, {Fuller}, {Richer}, {Harries},
  \& {Ladd}}]{hatchell-cores}
{Hatchell}, J., {Fuller}, G.~A., {Richer}, J.~S., {Harries}, T.~J., \& {Ladd},
  E.~F. 2007, \aap, 468, 1009

\bibitem[{{Hatchell} {et~al.}(2005){Hatchell}, {Richer}, {Fuller},
  {Qualtrough}, {Ladd}, \& {Chandler}}]{hatchell05}
{Hatchell}, J., {Richer}, J.~S., {Fuller}, G.~A., {Qualtrough}, C.~J., {Ladd},
  E.~F., \& {Chandler}, C.~J. 2005, \aap, 440, 151

\bibitem[{{Ho} {et~al.}(1979){Ho}, {Barrett}, {Myers}, {Matsakis}, {Chui},
  {Townes}, {Cheung}, \& {Yngvesson}}]{ho79}
{Ho}, P.~T.~P., {Barrett}, A.~H., {Myers}, P.~C., {Matsakis}, D.~N., {Chui},
  M.~F., {Townes}, C.~H., {Cheung}, A.~C., \& {Yngvesson}, K.~S. 1979, \apj,
  234, 912

\bibitem[{{Ho} \& {Townes}(1983)}]{nh3-araa}
{Ho}, P.~T.~P. \& {Townes}, C.~H. 1983, \araa, 21, 239

\bibitem[{{Jijina} {et~al.}(1999){Jijina}, {Myers}, \& {Adams}}]{jijina}
{Jijina}, J., {Myers}, P.~C., \& {Adams}, F.~C. 1999, \apjs, 125, 161

\bibitem[{{J{\o}rgensen} {et~al.}(2006){J{\o}rgensen}, {Harvey}, {Evans},
  {Huard}, {Allen}, {Porras}, {Blake}, {Bourke}, {Chapman}, {Cieza}, {Koerner},
  {Lai}, {Mundy}, {Myers}, {Padgett}, {Rebull}, {Sargent}, {Spiesman},
  {Stapelfeldt}, {van Dishoeck}, {Wahhaj}, \& {Young}}]{c2d-pers-irac}
{J{\o}rgensen}, J.~K., {Harvey}, P.~M., {Evans}, II, N.~J., {Huard}, T.~L.,
  {Allen}, L.~E., {Porras}, A., {Blake}, G.~A., {Bourke}, T.~L., {Chapman}, N.,
  {Cieza}, L., {Koerner}, D.~W., {Lai}, S.-P., {Mundy}, L.~G., {Myers}, P.~C.,
  {Padgett}, D.~L., {Rebull}, L., {Sargent}, A.~I., {Spiesman}, W.,
  {Stapelfeldt}, K.~R., {van Dishoeck}, E.~F., {Wahhaj}, Z., \& {Young}, K.~E.
  2006, \apj, 645, 1246

\bibitem[{{J{\o}rgensen} {et~al.}(2007){J{\o}rgensen}, {Johnstone}, {Kirk}, \&
  {Myers}}]{jorgensen-cores}
{J{\o}rgensen}, J.~K., {Johnstone}, D., {Kirk}, H., \& {Myers}, P.~C. 2007,
  \apj, 656, 293

\bibitem[{{Juan} {et~al.}(1993){Juan}, {Bachiller}, {Koempe}, \&
  {Martin-Pintado}}]{juan93}
{Juan}, J., {Bachiller}, R., {Koempe}, C., \& {Martin-Pintado}, J. 1993, \aap,
  270, 432

\bibitem[{{Kirk} {et~al.}(2006){Kirk}, {Johnstone}, \& {Di
  Francesco}}]{perseus-scuba}
{Kirk}, H., {Johnstone}, D., \& {Di Francesco}, J. 2006, \apj, 646, 1009

\bibitem[{{Ladd} {et~al.}(1994){Ladd}, {Myers}, \& {Goodman}}]{ladd-perseus}
{Ladd}, E.~F., {Myers}, P.~C., \& {Goodman}, A.~A. 1994, \apj, 433, 117

\bibitem[{{Langer} {et~al.}(1995){Langer}, {Velusamy}, {Kuiper}, {Levin},
  {Olsen}, \& {Migenes}}]{langer95}
{Langer}, W.~D., {Velusamy}, T., {Kuiper}, T.~B.~H., {Levin}, S., {Olsen}, E.,
  \& {Migenes}, V. 1995, \apj, 453, 293

\bibitem[{{Lovas} \& {Dragoset}(2003)}]{lovas-lines}
{Lovas}, F.~J. \& {Dragoset}, R. 2003, {Recommended rest frequencies for
  observed interstellar molecular microwave transitions} (Washington: National
  Bureau of Standards (NBS), 2003, Rev.~ed.)

\bibitem[{{Mauersberger} {et~al.}(1988){Mauersberger}, {Wilson}, \&
  {Henkel}}]{mauersberger88}
{Mauersberger}, R., {Wilson}, T.~L., \& {Henkel}, C. 1988, \aap, 201, 123

\bibitem[{{Motte} {et~al.}(1998){Motte}, {Andre}, \& {Neri}}]{motte-andre}
{Motte}, F., {Andre}, P., \& {Neri}, R. 1998, \aap, 336, 150

\bibitem[{{Ohishi} \& {Kaifu}(1998)}]{cc34s-freq}
{Ohishi}, M. \& {Kaifu}, N. 1998, in Chemistry and Physics of Molecules and
  Grains in Space. Faraday Discussions No. 109, 205--+

\bibitem[{{Pickett} {et~al.}(1998){Pickett}, {Poynter}, {Cohen}, {Delitsky},
  {Pearson}, \& {Muller}}]{jplcat}
{Pickett}, H.~M., {Poynter}, R.~L., {Cohen}, E.~A., {Delitsky}, M.~L.,
  {Pearson}, J.~C., \& {Muller}, H.~S.~P. 1998, J. Quant. Spectrosc. \& Rad.
  Transfer, 60, 883

\bibitem[{{Press} {et~al.}(1992){Press}, {Teukolsky}, {Vetterling}, \&
  {Flannery}}]{numrec}
{Press}, W.~H., {Teukolsky}, S.~A., {Vetterling}, W.~T., \& {Flannery}, B.~P.
  1992, {Numerical recipes in C. The art of scientific computing} (Cambridge:
  University Press, |c1992, 2nd ed.)

\bibitem[{{Rebull} {et~al.}(2007){Rebull}, {Stapelfeldt}, {Evans},
  {Joergensen}, {Harvey}, {Brooke}, {Bourke}, {Padgett}, {Chapman}, {Lai},
  {Spiesmann}, {Noreiga-Crespo}, {Merin}, {Huard}, {Allen}, {Blake}, {Jarrett},
  {Koerner}, {Mundy}, {Myers}, {Sargent}, {van Dishoeck}, {Wahhaj}, \&
  {Young}}]{c2d-pers-mips}
{Rebull}, L.~M., {Stapelfeldt}, K.~R., {Evans}, II, N.~J., {Joergensen}, J.~K.,
  {Harvey}, P.~M., {Brooke}, T.~Y., {Bourke}, T.~L., {Padgett}, D.~L.,
  {Chapman}, N.~L., {Lai}, S.~., {Spiesmann}, W.~J., {Noreiga-Crespo}, A.,
  {Merin}, B., {Huard}, T., {Allen}, L.~E., {Blake}, G.~A., {Jarrett}, T.,
  {Koerner}, D.~W., {Mundy}, L.~G., {Myers}, P.~C., {Sargent}, A.~I., {van
  Dishoeck}, E.~F., {Wahhaj}, Z., \& {Young}, K.~E. 2007, ArXiv Astrophysics
  e-prints

\bibitem[{{Ridge} {et~al.}(2006){Ridge}, {Di Francesco}, {Kirk}, {Li},
  {Goodman}, {Alves}, {Arce}, {Borkin}, {Caselli}, {Foster}, {Heyer},
  {Johnstone}, {Kosslyn}, {Lombardi}, {Pineda}, {Schnee}, \&
  {Tafalla}}]{complete-data}
{Ridge}, N.~A., {Di Francesco}, J., {Kirk}, H., {Li}, D., {Goodman}, A.~A.,
  {Alves}, J.~F., {Arce}, H.~G., {Borkin}, M.~A., {Caselli}, P., {Foster},
  J.~B., {Heyer}, M.~H., {Johnstone}, D., {Kosslyn}, D.~A., {Lombardi}, M.,
  {Pineda}, J.~E., {Schnee}, S.~L., \& {Tafalla}, M. 2006, \aj, 131, 2921

\bibitem[{{Rohlfs} \& {Wilson}(2004)}]{rohlfs-wilson}
{Rohlfs}, K. \& {Wilson}, T.~L. 2004, {Tools of radio astronomy} (Tools of
  radio astronomy, 4th rev.~and enl.~ed., by K.~Rohlfs and T.L.~Wilson.~
  Berlin: Springer, 2004)

\bibitem[{{Schnee} {et~al.}(in preparation){Schnee}, {Li}, \&
  {Goodman}}]{schnee-mips}
{Schnee}, S., {Li}, J., \& {Goodman}, A.~A. in preparation, \apj

\bibitem[{{Suzuki} {et~al.}(1992){Suzuki}, {Yamamoto}, {Ohishi}, {Kaifu},
  {Ishikawa}, {Hirahara}, \& {Takano}}]{suzuki92}
{Suzuki}, H., {Yamamoto}, S., {Ohishi}, M., {Kaifu}, N., {Ishikawa}, S.-I.,
  {Hirahara}, Y., \& {Takano}, S. 1992, \apj, 392, 551

\bibitem[{{Swade}(1989)}]{swade89}
{Swade}, D.~A. 1989, \apj, 345, 828

\bibitem[{{Swift} {et~al.}(2005){Swift}, {Welch}, \& {Di
  Francesco}}]{swift-nh3}
{Swift}, J.~J., {Welch}, W.~J., \& {Di Francesco}, J. 2005, \apj, 620, 823

\bibitem[{{Tafalla} {et~al.}(2004){Tafalla}, {Myers}, {Caselli}, \&
  {Walmsley}}]{tafalla04}
{Tafalla}, M., {Myers}, P.~C., {Caselli}, P., \& {Walmsley}, C.~M. 2004, \aap,
  416, 191

\bibitem[{{Testi} \& {Sargent}(1998)}]{testi-sargent}
{Testi}, L. \& {Sargent}, A.~I. 1998, \apjl, 508, L91

\bibitem[{{Wilson} \& {Rood}(1994)}]{ism-abund}
{Wilson}, T.~L. \& {Rood}, R. 1994, \araa, 32, 191

\bibitem[{{Yamamoto} {et~al.}(1990){Yamamoto}, {Saito}, {Kawaguchi}, {Chikada},
  {Suzuki}, {Kaifu}, {Ishikawa}, \& {Ohishi}}]{ccs-freq}
{Yamamoto}, S., {Saito}, S., {Kawaguchi}, K., {Chikada}, Y., {Suzuki}, H.,
  {Kaifu}, N., {Ishikawa}, S.-I., \& {Ohishi}, M. 1990, \apj, 361, 318

\end{thebibliography}

\stepcounter{figure}
\begin{figure}
\figurenum{\arabic{figure}a}
\epsscale{0.8}
\plotone{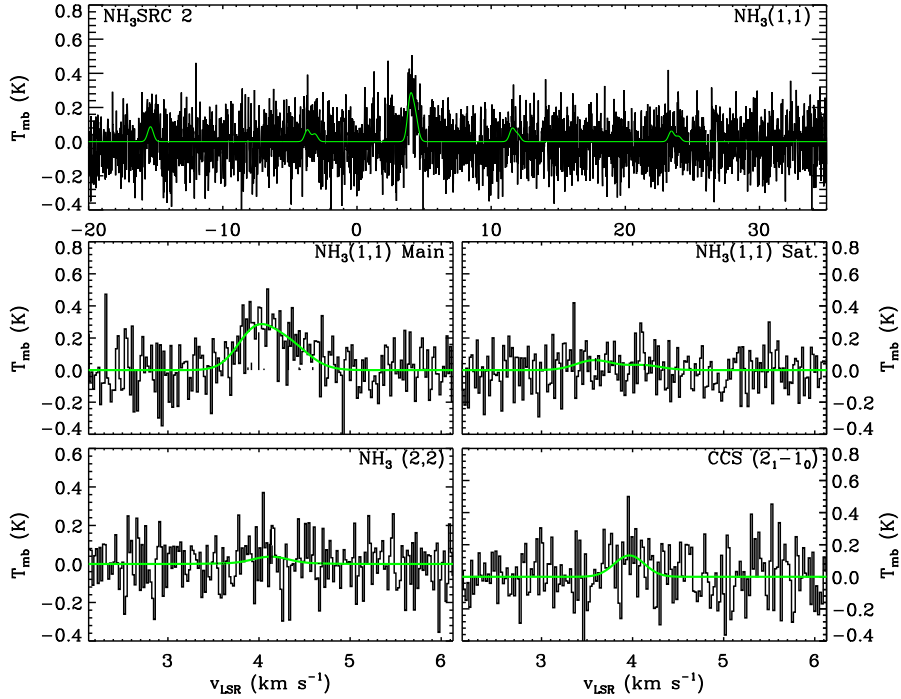}
\caption{\label{fullspex} ({\it top}) The full NH$_3$(1,1) spectrum of NH$_3$SRC 1.  The spectrum is shown in black, the best fitting model described in \S\ref{params} is shown in green.  Where discernible, additional velocity components are plotted in blue.  ({\it middle,left}) The central component of the NH$_3$(1,1) line and best fitting model.  The velocities and relative amplitudes of the hyperfine components are indicated with vertical lines.  ({\it middle, right}) The NH$_3$(1,1) high-velocity hyperfine satellite at $\Delta v=19.85\mbox{ km s}^{-1}$ and best fitting model.  Hyperfine components are indicated with vertical lines.  ({\it bottom, left}) The main component of NH$_3$(2,2) line and best fitting model. Hyperfine components are indicated with vertical lines. ({\it bottom, right})  The C$_{2}$S ($2_1\to 1_0$) line and best fitting model.}
\epsscale{1.0}
\end{figure}

\begin{figure}
\figurenum{\arabic{figure}b}
\epsscale{0.8}
\plotone{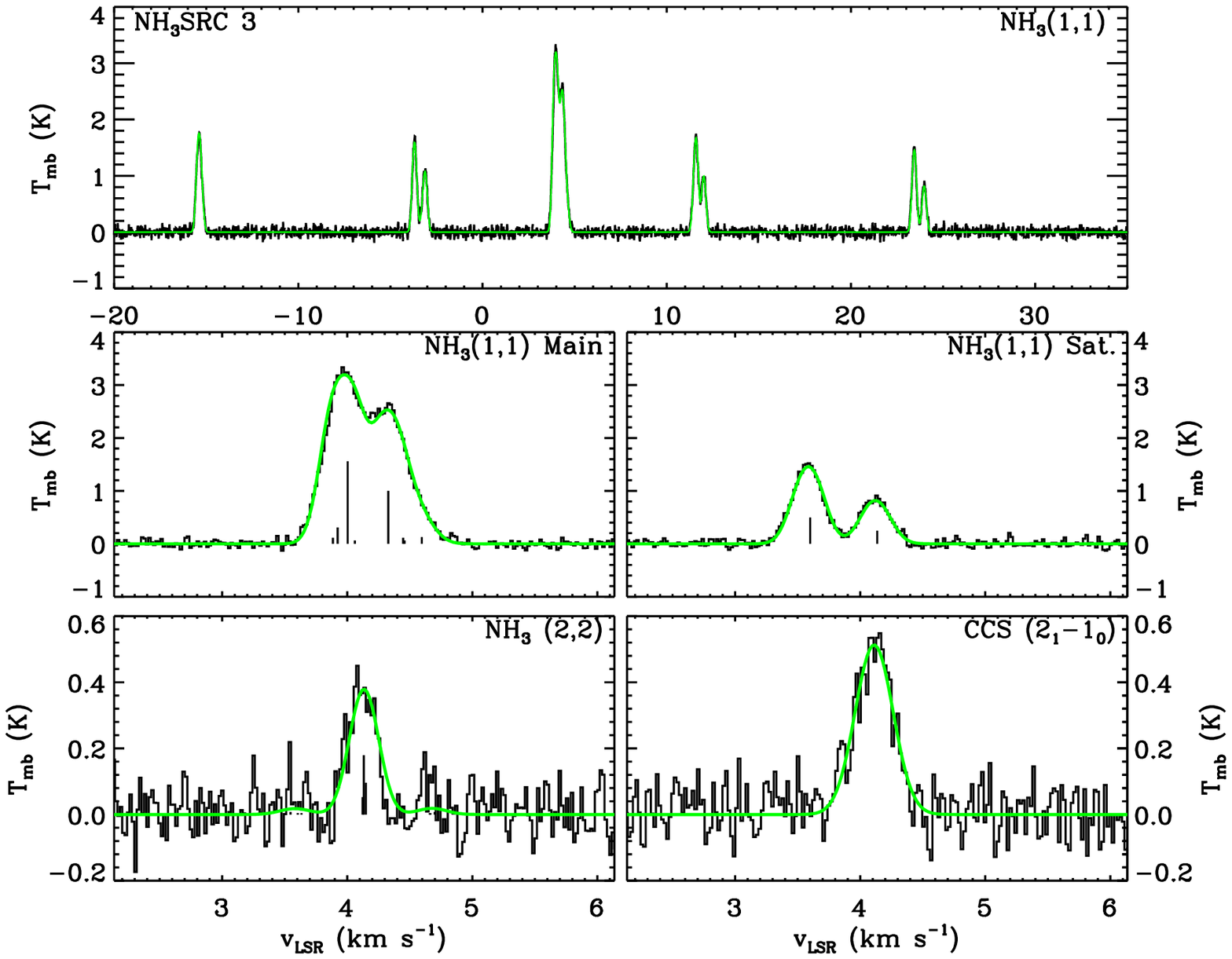}
\caption{As Figure \arabic{figure}a but for NH$_3$SRC 3.}
\epsscale{1.0}
\end{figure}
 
\begin{figure}
\figurenum{\arabic{figure}c}
\epsscale{0.8}
\plotone{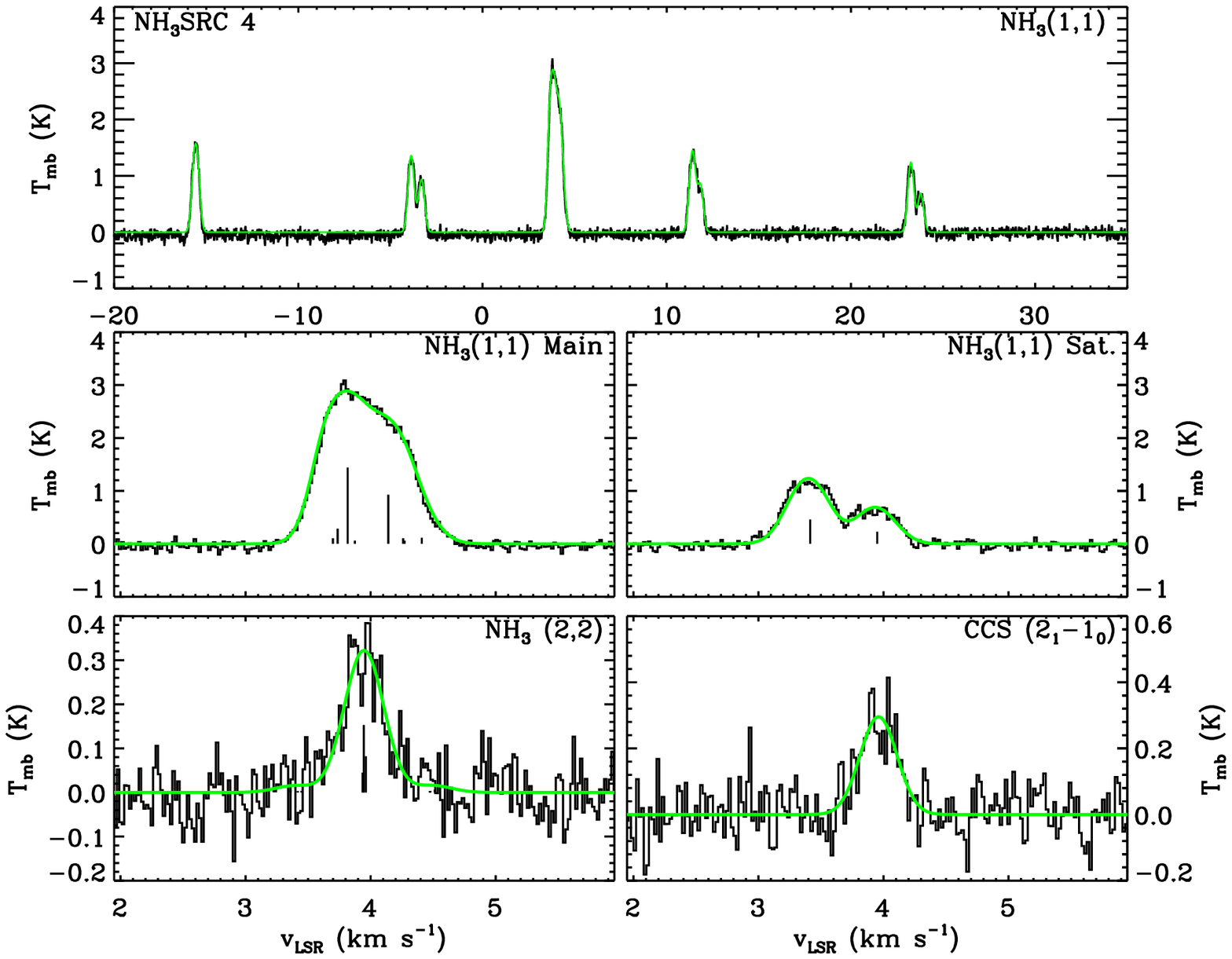}
\caption{As Figure \arabic{figure}a but for NH$_3$SRC 4.}
\epsscale{1.0}
\end{figure}

\begin{figure}
\figurenum{\arabic{figure}d}
\epsscale{0.8}
\plotone{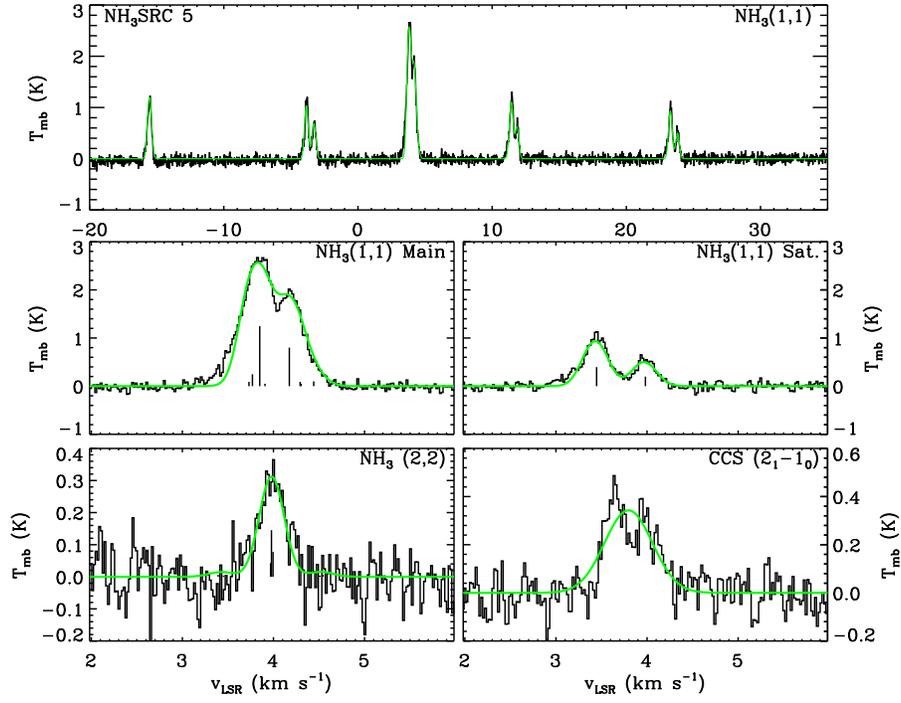}
\caption{As Figure \arabic{figure}a but for NH$_3$SRC 5.}
\epsscale{1.0}
\end{figure}

\clearpage
\LongTables
%\begin{landscape}

\begin{deluxetable}{ccccccccrrr}
\tablecaption{\label{obstable}Summary of Observations}
\tabletypesize{\footnotesize}
\tablewidth{0pt}
\tablehead{
\colhead{NH3SRC} & \colhead{Origin} & \colhead{Region} & \colhead{Position} &
\colhead{Bolocam} & \colhead{SCUBA} &
\colhead{Int. Time} & \colhead{$\sigma_{rms}$} &
\colhead{$W[\mathrm{NH}_{3}(1,1)]$} &
\colhead{$W[\mathrm{NH}_{3}(2,2)]$} &
\colhead{$W[\mathrm{C_2S}]$}\\
\colhead{} & \colhead{} & \colhead{} &
\colhead{($\alpha_{2000}$,~$\delta_{2000}$)} &
\colhead{Name} & \colhead{Name} &
\colhead{(min.)} & \colhead{(mK)} &
\colhead{(K~km~s$^{-1}$)} &\colhead{(K~km~s$^{-1}$)}
&\colhead{(K~km~s$^{-1}$)} \\
\colhead{(1)} & \colhead{(2)} & \colhead{(3)} & \colhead{(4)} &
\colhead{(5)} & \colhead{(6)} & \colhead{(7)} & \colhead{(8)} &
\colhead{(9)} & \colhead{(10)} & \colhead{(11)}
}
\startdata
1 & D & \nodata & 03:22:18.9~+30:53:14 & \nodata & N/A & 5 & 120 & 0.07(8) & 0.03(3) & 0.00(4) \\ 
2 & D & L1455/L1448 & 03:25:00.3~+30:44:10 & \nodata & \nodata & 5 & 128 & 0.46(9) & 0.02(4) & 0.05(5) \\ 
3 & B & L1455/L1448 & 03:25:07.8~+30:24:22 & 1 & \nodata & 15 & 63 & 5.25(4) & 0.15(2) & 0.22(2) \\ 
4 & B & L1455/L1448 & 03:25:09.7~+30:23:53 & 2 & \nodata & 15 & 60 & 5.87(4) & 0.12(2) & 0.09(2) \\ 
5 & B & L1455/L1448 & 03:25:10.1~+30:44:41 & 3 & \nodata & 15 & 63 & 4.08(4) & 0.10(2) & 0.23(2) \\ 
6 & B & L1455/L1448 & 03:25:17.1~+30:18:53 & 4 & N/A & 20 & 54 & 2.21(4) & 0.04(1) & 0.36(2) \\ 
7 & B & L1455/L1448 & 03:25:22.3~+30:45:09 & 5 & 032537+30451 & 15 & 60 & 10.09(4) & 0.67(2) & 0.20(2) \\ 
8 & S & L1455/L1448 & 03:25:26.2~+30:45:05 & \nodata & 032543+30450 & 10 & 87 & 12.36(6) & 0.79(2) & 0.34(3) \\ 
9 & B & L1455/L1448 & 03:25:26.9~+30:21:53 & 6 & N/A & 15 & 61 & 4.43(4) & 0.10(2) & 0.23(2) \\ 
10 & D & L1455/L1448 & 03:25:32.3~+30:46:00 & \nodata & \nodata & 10 & 74 & 2.93(5) & 0.10(2) & 0.06(3) \\ 
11 & B & L1455/L1448 & 03:25:35.5~+30:13:06 & 7 & N/A & 15 & 58 & 0.94(4) & 0.04(2) & 0.36(2) \\ 
12 & B & L1455/L1448 & 03:25:36.2~+30:45:11 & 8 & 032560+30453 & 20 & 54 & 19.97(4) & 1.65(1) & 0.24(2) \\ 
13 & B & L1455/L1448 & 03:25:37.2~+30:09:55 & 9 & N/A & 20 & 44 & 0.61(3) & 0.04(1) & 0.12(2) \\ 
14 & B & L1455/L1448 & 03:25:38.6~+30:43:59 & 10 & 032564+30440 & 10 & 62 & 13.29(4) & 1.29(2) & 0.38(2) \\ 
15 & B & L1455/L1448 & 03:25:46.1~+30:44:11 & 11 & \nodata & 10 & 61 & 5.62(4) & 0.26(2) & 0.16(2) \\ 
16 & B & L1455/L1448 & 03:25:47.5~+30:12:26 & 12 & N/A & 20 & 41 & 0.71(3) & 0.02(1) & 0.36(2) \\ 
17 & B & L1455/L1448 & 03:25:48.8~+30:42:24 & 13 & 032581+30423 & 15 & 40 & 11.49(3) & 0.34(1) & 0.26(2) \\ 
18 & B & L1455/L1448 & 03:25:50.6~+30:42:02 & 14 & \nodata & 10 & 66 & 10.32(5) & 0.27(2) & 0.25(3) \\ 
19 & B & L1455/L1448 & 03:25:55.1~+30:41:26 & 15 & \nodata & 25 & 42 & 1.63(3) & 0.03(1) & 0.20(2) \\ 
20 & B & L1455/L1448 & 03:25:56.4~+30:40:43 & 16 & \nodata & 25 & 39 & 0.80(3) & 0.02(1) & 0.16(1) \\ 
21 & B & L1455/L1448 & 03:25:58.5~+30:37:14 & 17 & \nodata & 25 & 39 & 0.89(3) & 0.02(1) & 0.07(2) \\ 
22 & B & L1455/L1448 & 03:26:37.0~+30:15:23 & 18 & 032662+30153 & 10 & 60 & 4.89(4) & 0.30(2) & 0.26(2) \\ 
23 & D & \nodata & 03:26:39.7~+31:28:21 & N/A & N/A & 5 & 94 & 0.30(7) & 0.02(3) & 0.07(4) \\ 
24 & B & L1455/L1448 & 03:27:02.1~+30:15:08 & 19 & \nodata & 15 & 44 & 1.74(3) & 0.05(1) & 0.21(2) \\ 
25 & D & \nodata & 03:27:14.3~+31:32:49 & \nodata & N/A & 5 & 138 & 0.7(1) & 0.06(4) & -0.04(5) \\ 
26 & D & L1455/L1448 & 03:27:20.2~+30:04:26 & \nodata & N/A & 10 & 77 & 0.95(5) & 0.04(2) & 0.00(3) \\ 
27 & D & L1455/L1448 & 03:27:20.5~+30:00:42 & N/A & N/A & 5 & 108 & 0.75(7) & 0.08(3) & 0.18(4) \\ 
28 & D & \nodata & 03:27:26.4~+29:51:08 & N/A & N/A & 10 & 122 & 0.74(8) & 0.01(4) & 0.19(5) \\ 
29 & B & L1455/L1448 & 03:27:28.9~+30:15:04 & 20 & \nodata & 20 & 54 & 4.51(4) & 0.16(2) & 0.36(2) \\ 
30 & L & L1455/L1448 & 03:27:34.4~+30:09:22 & \nodata & \nodata & 5 & 87 & 1.54(6) & 0.02(2) & 0.13(3) \\ 
31 & B & L1455/L1448 & 03:27:37.7~+30:14:00 & 21 & 032763+30139 & 30 & 38 & 6.04(3) & 0.31(1) & 0.15(1) \\ 
32 & B & L1455/L1448 & 03:27:39.3~+30:12:59 & 22 & 032765+30130 & 20 & 45 & 10.56(3) & 0.85(1) & 0.20(2) \\ 
33 & S & L1455/L1448 & 03:27:40.0~+30:12:13 & \nodata & 032766+30122 & 5 & 136 & 11.96(9) & 0.55(4) & 0.30(6) \\ 
34 & B & L1455/L1448 & 03:27:41.9~+30:12:30 & 23 & 032771+30125 & 20 & 41 & 10.78(3) & 0.61(1) & 0.25(2) \\ 
35 & B & L1455/L1448 & 03:27:47.9~+30:12:02 & 24 & 032780+30121 & 30 & 39 & 6.36(3) & 0.38(1) & 0.16(2) \\ 
36 & L & L1455/L1448 & 03:27:55.9~+30:06:18 & \nodata & N/A & 5 & 84 & 4.96(6) & 0.12(2) & 0.17(3) \\ 
37 & L & L1455/L1448 & 03:28:00.7~+30:08:20 & \nodata & \nodata & 5 & 132 & 4.01(9) & 0.18(4) & 0.25(5) \\ 
38 & L & L1455/L1448 & 03:28:05.5~+30:06:19 & \nodata & N/A & 5 & 80 & 5.00(6) & 0.14(2) & 0.14(3) \\ 
39 & W & NGC1333 & 03:28:30.0~+30:55:29 & \nodata & \nodata & 5 & 131 & 0.09(9) & -0.01(4) & 0.05(6) \\ 
40 & B & NGC1333 & 03:28:32.2~+31:11:09 & 25 & \nodata & 15 & 38 & 5.36(3) & 0.25(1) & 0.05(2) \\ 
41 & B & NGC1333 & 03:28:32.4~+31:04:43 & 26 & \nodata & 10 & 54 & 6.36(4) & 0.29(2) & 0.10(2) \\ 
42 & B & L1455/L1448 & 03:28:33.4~+30:19:35 & 27 & \nodata & 15 & 68 & 2.24(5) & 0.12(2) & 0.66(2) \\ 
43 & B & NGC1333 & 03:28:34.1~+31:07:01 & 28 & \nodata & 15 & 54 & 3.01(4) & 0.12(2) & 0.06(2) \\ 
44 & B & NGC1333 & 03:28:36.3~+31:13:27 & 29 & 032861+31134 & 15 & 41 & 5.26(3) & 0.34(1) & 0.02(2) \\ 
45 & W & NGC1333 & 03:28:36.8~+31:00:14 & \nodata & \nodata & 5 & 142 & 0.00(1) & 0.00(4) & 0.03(6) \\ 
46 & B & NGC1333 & 03:28:39.1~+31:06:00 & 30 & 032865+31060 & 10 & 57 & 8.85(4) & 0.42(2) & 0.24(2) \\ 
47 & S & NGC1333 & 03:28:39.5~+31:18:35 & \nodata & 032865+31185 & 41 & 34 & 12.60(2) & 0.847(9) & 0.18(1) \\ 
48 & S & NGC1333 & 03:28:40.3~+31:17:56 & 31 & 032866+31179 & 5 & 170 & 16.2(1) & 1.09(4) & 0.10(7) \\ 
49 & B & L1455/L1448 & 03:28:41.7~+30:31:12 & 32 & \nodata & 15 & 41 & 0.94(3) & -0.02(1) & 0.37(2) \\ 
50 & B & NGC1333 & 03:28:42.6~+31:06:13 & 33 & \nodata & 10 & 57 & 8.26(4) & 0.37(2) & 0.14(2) \\ 
51 & B & NGC1333 & 03:28:46.0~+31:15:19 & 34 & \nodata & 10 & 65 & 8.78(4) & 0.42(2) & 0.08(3) \\ 
52 & B & NGC1333 & 03:28:48.5~+31:16:03 & 35 & \nodata & 10 & 66 & 7.22(5) & 0.39(2) & 0.06(3) \\ 
53 & B & L1455/L1448 & 03:28:48.8~+30:43:25 & 36 & \nodata & 20 & 40 & 0.63(3) & 0.04(1) & 0.10(2) \\ 
54 & D & NGC1333 & 03:28:49.6~+31:30:01 & \nodata & \nodata & 10 & 75 & 0.20(5) & -0.02(2) & 0.00(3) \\ 
55 & D & L1455/L1448 & 03:28:51.4~+30:32:58 & \nodata & \nodata & 10 & 113 & 0.77(8) & 0.02(3) & 0.05(4) \\ 
56 & B & NGC1333 & 03:28:52.2~+31:18:08 & 37 & \nodata & 10 & 76 & 6.12(5) & 0.52(2) & 0.04(3) \\ 
57 & D & NGC1333 & 03:28:55.2~+31:20:26 & \nodata & \nodata & 10 & 82 & 2.16(6) & 0.24(2) & -0.04(3) \\ 
58 & B & NGC1333 & 03:28:55.3~+31:14:33 & 38 & 032891+31145 & 10 & 63 & 11.13(4) & 1.44(2) & 0.17(3) \\ 
59 & B & NGC1333 & 03:28:55.4~+31:19:19 & 39 & \nodata & 10 & 71 & 8.00(5) & 0.87(2) & 0.02(3) \\ 
60 & D & L1455/L1448 & 03:28:56.2~+30:03:42 & N/A & N/A & 10 & 97 & 0.51(7) & 0.02(3) & 0.13(4) \\ 
61 & D & NGC1333 & 03:28:57.5~+31:23:06 & \nodata & \nodata & 10 & 80 & 0.66(6) & 0.07(2) & 0.03(3) \\ 
62 & D & L1455/L1448 & 03:28:58.1~+30:45:12 & \nodata & \nodata & 10 & 104 & 0.53(7) & 0.04(3) & 0.01(4) \\ 
63 & D & NGC1333 & 03:28:58.6~+31:09:10 & \nodata & \nodata & 5 & 115 & 0.56(8) & 0.03(3) & -0.08(4) \\ 
64 & B & NGC1333 & 03:28:59.6~+31:21:38 & 40 & 032899+31215 & 10 & 69 & 8.47(5) & 0.85(2) & 0.08(3) \\ 
65 & B & NGC1333 & 03:29:00.6~+31:11:59 & 41 & 032900+31119 & 10 & 61 & 7.87(4) & 0.54(2) & 0.03(2) \\ 
66 & B & NGC1333 & 03:29:01.4~+31:20:34 & 42 & 032901+31204 & 10 & 86 & 11.46(6) & 1.55(2) & 0.03(3) \\ 
67 & S & NGC1333 & 03:29:03.2~+31:15:59 & 43 & 032905+31159 & 10 & 113 & 12.93(8) & 1.71(3) & 0.05(4) \\ 
68 & S & NGC1333 & 03:29:03.4~+31:14:58 & \nodata & 032905+31149 & 10 & 80 & 16.77(6) & 2.27(2) & 0.09(3) \\ 
69 & B & NGC1333 & 03:29:04.5~+31:18:43 & 44 & \nodata & 10 & 69 & 6.41(5) & 0.54(2) & 0.06(3) \\ 
70 & S & NGC1333 & 03:29:06.9~+31:15:44 & \nodata & 032910+31156 & 15 & 101 & 13.10(7) & 1.45(3) & 0.17(4) \\ 
71 & S & NGC1333 & 03:29:07.5~+31:21:54 & \nodata & 032912+31218 & 10 & 57 & 0.75(4) & 0.19(2) & -0.02(2) \\ 
72 & B & NGC1333 & 03:29:07.8~+31:17:19 & 45 & 032911+31173 & 10 & 72 & 7.35(5) & 0.55(2) & 0.03(3) \\ 
73 & S & NGC1333 & 03:29:08.9~+31:15:12 & 46 & 032914+31152 & 10 & 56 & 18.40(4) & 1.31(2) & 0.20(2) \\ 
74 & D & L1455/L1448 & 03:29:09.6~+30:21:18 & \nodata & \nodata & 5 & 112 & 0.21(8) & 0.00(3) & 0.00(4) \\ 
75 & S & NGC1333 & 03:29:10.3~+31:13:35 & 48 & 032916+31135 & 15 & 58 & 11.28(4) & 1.25(2) & 0.13(2) \\ 
76 & S & NGC1333 & 03:29:10.3~+31:21:44 & \nodata & 032917+31217 & 10 & 181 & 0.9(1) & 0.15(5) & 0.00(7) \\ 
77 & B & NGC1333 & 03:29:11.4~+31:18:26 & 49 & 032917+31184 & 10 & 65 & 10.10(5) & 1.02(2) & 0.06(3) \\ 
78 & S & NGC1333 & 03:29:11.4~+31:13:07 & \nodata & 032919+31131 & 10 & 264 & 7.4(2) & 0.67(7) & 0.1(1) \\ 
79 & B & NGC1333 & 03:29:14.9~+31:20:27 & 50 & 032925+31205 & 15 & 52 & 2.08(4) & 0.20(1) & -0.01(2) \\ 
80 & B & NGC1333 & 03:29:17.0~+31:12:26 & 51 & \nodata & 30 & 54 & 6.02(4) & 0.28(1) & 0.02(2) \\ 
81 & B & NGC1333 & 03:29:17.2~+31:27:40 & 52 & 032928+31278 & 10 & 69 & 4.07(5) & 0.25(2) & 0.01(3) \\ 
82 & B & NGC1333 & 03:29:18.5~+31:25:13 & 53 & 032930+31251 & 15 & 54 & 4.62(4) & 0.32(1) & 0.03(2) \\ 
83 & B & NGC1333 & 03:29:19.1~+31:11:32 & 55 & \nodata & 35 & 50 & 5.56(3) & 0.21(1) & 0.06(2) \\ 
84 & B & NGC1333 & 03:29:19.2~+31:23:28 & 54 & \nodata & 20 & 72 & 2.44(5) & 0.18(2) & -0.09(3) \\ 
85 & D & NGC1333 & 03:29:20.5~+31:19:30 & \nodata & \nodata & 10 & 73 & 0.55(5) & 0.05(2) & 0.04(3) \\ 
86 & B & NGC1333 & 03:29:22.5~+31:36:24 & 56 & \nodata & 20 & 67 & 2.93(5) & 0.04(2) & 0.01(3) \\ 
87 & B & NGC1333 & 03:29:22.9~+31:33:16 & 57 & 032939+31333 & 25 & 65 & 6.81(5) & 0.31(2) & 0.02(3) \\ 
88 & B & NGC1333 & 03:29:25.8~+31:28:17 & 58 & 032942+31283 & 10 & 63 & 7.67(4) & 0.28(2) & 0.05(3) \\ 
89 & B & NGC1333 & 03:29:51.5~+31:39:12 & 59 & 032986+31391 & 20 & 70 & 6.80(5) & 0.29(2) & 0.08(3) \\ 
90 & D & NGC1333 & 03:30:13.6~+31:44:38 & \nodata & N/A & 10 & 121 & 0.36(8) & 0.03(3) & -0.07(4) \\ 
91 & B & B1-W & 03:30:15.1~+30:23:39 & 60 & \nodata & 10 & 69 & 7.44(5) & 0.37(2) & 0.03(2) \\ 
92 & D & B1-W & 03:30:23.1~+30:31:09 & \nodata & \nodata & 5 & 87 & 0.22(6) & 0.02(2) & 0.02(3) \\ 
93 & B & B1-W & 03:30:24.1~+30:27:39 & 61 & \nodata & 15 & 43 & 2.32(3) & 0.07(1) & 0.27(2) \\ 
94 & D & B1-W & 03:30:25.7~+30:31:10 & \nodata & \nodata & 5 & 109 & -0.09(8) & -0.02(3) & 0.00(4) \\ 
95 & B & B1-W & 03:30:32.0~+30:26:19 & 62 & \nodata & 10 & 56 & 10.88(4) & 0.49(2) & 0.21(2) \\ 
96 & B & B1-W & 03:30:45.6~+30:52:36 & 63 & \nodata & 20 & 60 & 2.70(4) & 0.07(2) & 0.32(2) \\ 
97 & B & B1-W & 03:30:50.5~+30:49:17 & 64 & \nodata & 23 & 62 & 2.47(4) & 0.04(2) & 0.13(2) \\ 
98 & D & B1-W & 03:31:14.4~+30:44:03 & \nodata & \nodata & 10 & 127 & 1.15(9) & 0.17(4) & 0.01(5) \\ 
99 & B & B1-W & 03:31:20.0~+30:45:30 & 65 & 033134+30454 & 15 & 40 & 8.86(3) & 0.53(1) & 0.03(2) \\ 
100 & D & \nodata & 03:31:58.4~+30:02:04 & N/A & N/A & 10 & 96 & 0.12(7) & 0.04(3) & 0.01(3) \\ 
101 & D & B1 & 03:32:10.1~+31:19:54 & \nodata & \nodata & 5 & 130 & 0.62(9) & -0.02(4) & 0.03(5) \\ 
102 & W & B1 & 03:32:17.5~+30:53:58 & \nodata & \nodata & 10 & 145 & 0.7(1) & 0.03(4) & 0.21(6) \\ 
103 & B & B1 & 03:32:17.5~+30:49:49 & 66 & 033229+30497 & 15 & 63 & 11.28(4) & 0.72(2) & 0.24(3) \\ 
104 & B & B1 & 03:32:26.9~+30:59:11 & 67 & \nodata & 15 & 40 & 6.54(3) & 0.28(1) & 0.64(1) \\ 
105 & B & B1 & 03:32:28.1~+31:02:19 & 68 & \nodata & 15 & 55 & 3.66(4) & 0.11(2) & 0.14(2) \\ 
106 & W & B1 & 03:32:28.6~+30:53:51 & \nodata & \nodata & 10 & 155 & 0.6(1) & -0.03(4) & 0.05(6) \\ 
107 & B & B1 & 03:32:39.3~+30:57:29 & 69 & \nodata & 32 & 43 & 1.42(3) & 0.03(1) & 0.18(2) \\ 
108 & B & B1 & 03:32:44.1~+31:00:01 & 70 & \nodata & 10 & 60 & 8.72(4) & 0.35(2) & 0.28(2) \\ 
109 & B & B1 & 03:32:51.3~+31:01:48 & 71 & \nodata & 20 & 42 & 1.77(3) & 0.07(1) & 0.22(2) \\ 
110 & D & B1 & 03:32:54.8~+31:19:23 & \nodata & \nodata & 10 & 75 & 0.76(5) & 0.04(2) & 0.06(3) \\ 
111 & B & B1 & 03:32:57.0~+31:03:21 & 72 & \nodata & 15 & 42 & 5.89(3) & 0.24(1) & 0.19(2) \\ 
112 & B & B1 & 03:33:00.1~+31:20:45 & 73 & \nodata & 20 & 45 & 1.74(3) & 0.10(1) & 0.19(2) \\ 
113 & B & B1 & 03:33:02.0~+31:04:33 & 74 & 033303+31044 & 10 & 57 & 7.87(4) & 0.29(2) & 0.39(2) \\ 
114 & B & B1 & 03:33:04.3~+31:04:57 & 75 & \nodata & 15 & 49 & 11.78(3) & 0.48(1) & 0.41(2) \\ 
115 & W & B1 & 03:33:06.3~+31:06:26 & \nodata & \nodata & 5 & 154 & 4.4(1) & 0.21(4) & 0.16(6) \\ 
116 & B & B1 & 03:33:11.5~+31:17:23 & 77 & \nodata & 15 & 70 & 1.07(5) & 0.03(2) & 0.16(3) \\ 
117 & B & B1 & 03:33:11.6~+31:21:33 & 76 & \nodata & 15 & 69 & 0.83(5) & 0.03(2) & 0.07(3) \\ 
118 & B & B1 & 03:33:13.3~+31:19:51 & 78 & 033322+31199 & 10 & 69 & 9.78(5) & 0.36(2) & 0.11(3) \\ 
119 & B & B1 & 03:33:15.1~+31:07:04 & 79 & 033326+31069 & 15 & 41 & 15.75(3) & 1.03(1) & 0.44(1) \\ 
120 & D & B1 & 03:33:17.6~+31:17:19 & \nodata & \nodata & 5 & 107 & 0.55(7) & 0.00(3) & -0.08(4) \\ 
121 & B & B1 & 03:33:17.9~+31:09:30 & 80 & 033329+31095 & 20 & 40 & 16.40(3) & 1.23(1) & 0.41(2) \\ 
122 & D & B1 & 03:33:19.8~+31:22:41 & \nodata & \nodata & 10 & 75 & 0.70(5) & 0.03(2) & 0.01(3) \\ 
123 & B & B1 & 03:33:20.5~+31:07:37 & 81 & 033335+31075 & 15 & 45 & 18.75(3) & 1.25(1) & 0.43(2) \\ 
124 & B & B1 & 03:33:25.2~+31:05:35 & 82 & \nodata & 10 & 56 & 6.69(4) & 0.27(2) & 0.14(2) \\ 
125 & B & B1 & 03:33:25.4~+31:20:05 & 83 & \nodata & 20 & 47 & 2.05(3) & 0.14(1) & 0.21(2) \\ 
126 & B & B1 & 03:33:27.1~+31:06:56 & 84 & \nodata & 15 & 57 & 4.13(4) & 0.23(2) & 0.27(2) \\ 
127 & B & B1 & 03:33:31.8~+31:20:02 & 85 & \nodata & 20 & 56 & 3.58(4) & 0.14(2) & 0.07(2) \\ 
128 & B & B1 & 03:33:51.2~+31:12:38 & 86 & \nodata & 20 & 41 & 1.16(3) & 0.03(1) & 0.01(2) \\ 
129 & D & B1 & 03:33:52.5~+31:22:37 & \nodata & \nodata & 10 & 120 & 0.99(8) & 0.08(3) & -0.02(5) \\ 
130 & D & B1-E & 03:34:59.9~+31:15:17 & \nodata & \nodata & 7 & 182 & 0.3(1) & -0.07(5) & -0.10(7) \\ 
131 & D & B1-E & 03:35:02.2~+30:55:06 & \nodata & \nodata & 5 & 117 & -0.14(8) & 0.00(3) & 0.00(4) \\ 
132 & B & B1-E & 03:35:21.5~+31:06:56 & 87 & \nodata & 15 & 59 & 0.73(4) & 0.06(2) & 0.11(3) \\ 
133 & D & B1-E & 03:35:42.6~+31:09:53 & \nodata & \nodata & 5 & 116 & 0.30(9) & -0.01(4) & 0.04(5) \\ 
134 & W & B1-E & 03:35:56.5~+31:15:21 & \nodata & \nodata & 5 & 143 & 0.4(1) & 0.00(4) & 0.03(6) \\ 
135 & D & B1-E & 03:36:44.6~+31:13:42 & \nodata & \nodata & 5 & 128 & 0.19(9) & 0.04(4) & 0.00(5) \\ 
136 & W & \nodata & 03:37:07.5~+31:33:25 & \nodata & N/A & 5 & 146 & -0.2(1) & -0.03(4) & -0.14(5) \\ 
137 & W & \nodata & 03:37:14.5~+31:24:16 & \nodata & \nodata & 5 & 146 & 0.0(1) & 0.06(4) & 0.03(6) \\ 
138 & D & \nodata & 03:38:09.0~+30:46:28 & \nodata & N/A & 5 & 119 & 0.21(8) & 0.02(3) & -0.01(4) \\ 
139 & W & \nodata & 03:38:15.1~+31:19:45 & \nodata & \nodata & 10 & 53 & 0.43(4) & 0.10(2) & 0.02(2) \\ 
140 & D & \nodata & 03:39:00.6~+30:41:07 & N/A & N/A & 5 & 120 & 0.56(8) & 0.02(3) & -0.03(4) \\ 
141 & B & IC348 & 03:40:14.5~+32:01:30 & 88 & N/A & 10 & 140 & 1.2(1) & 0.08(4) & 0.05(6) \\ 
142 & B & IC348 & 03:40:49.5~+31:48:35 & 89 & \nodata & 10 & 61 & 2.17(4) & 0.14(2) & 0.14(2) \\ 
143 & B & IC348 & 03:41:09.3~+31:44:33 & 90 & \nodata & 20 & 71 & 0.02(5) & -0.06(2) & -0.04(3) \\ 
144 & B & IC348 & 03:41:19.9~+31:47:28 & 91 & \nodata & 10 & 51 & 0.47(4) & 0.06(1) & 0.00(2) \\ 
145 & B & IC348 & 03:41:40.2~+31:58:05 & 92 & \nodata & 5 & 156 & 4.8(1) & 0.19(4) & 0.01(6) \\ 
146 & B & IC348 & 03:41:45.2~+31:48:09 & 93 & \nodata & 10 & 51 & 0.59(4) & 0.07(1) & 0.01(2) \\ 
147 & B & IC348 & 03:41:46.0~+31:57:22 & 94 & \nodata & 5 & 70 & 5.90(5) & 0.24(2) & 0.01(2) \\ 
148 & W & IC348 & 03:41:58.3~+31:58:36 & \nodata & \nodata & 5 & 223 & 1.3(2) & 0.11(6) & 0.14(9) \\ 
149 & W & IC348 & 03:42:09.0~+31:46:50 & \nodata & \nodata & 5 & 204 & 0.3(1) & -0.01(6) & 0.03(8) \\ 
150 & B & IC348 & 03:42:20.3~+31:44:51 & 95 & \nodata & 15 & 51 & 0.43(4) & 0.04(1) & 0.00(2) \\ 
151 & W & IC348 & 03:42:24.0~+31:45:43 & \nodata & \nodata & 15 & 56 & 0.59(4) & 0.03(2) & 0.04(2) \\ 
152 & B & IC348 & 03:42:47.2~+31:58:41 & 96 & \nodata & 5 & 135 & 0.95(9) & 0.17(4) & 0.16(6) \\ 
153 & B & IC348 & 03:42:52.5~+31:58:11 & 97 & \nodata & 5 & 168 & 0.9(1) & 0.10(5) & 0.03(7) \\ 
154 & B & IC348 & 03:42:57.3~+31:57:48 & 98 & \nodata & 10 & 150 & 1.0(1) & 0.12(4) & -0.07(6) \\ 
155 & W & IC348 & 03:43:29.6~+31:55:22 & \nodata & \nodata & 5 & 204 & 0.5(1) & 0.01(6) & 0.03(8) \\ 
156 & B & IC348 & 03:43:38.1~+32:03:10 & 99 & 034363+32032 & 5 & 150 & 4.3(1) & 0.35(4) & 0.06(6) \\ 
157 & S & IC348 & 03:43:44.0~+32:02:52 & \nodata & 034373+32028 & 15 & 44 & 3.89(3) & 0.27(1) & 0.06(2) \\ 
158 & B & IC348 & 03:43:45.5~+32:01:44 & 101 & \nodata & 15 & 52 & 1.50(4) & 0.13(1) & 0.06(2) \\ 
159 & S & IC348 & 03:43:45.8~+32:03:11 & \nodata & 034376+32031 & 10 & 54 & 5.63(4) & 0.35(2) & 0.02(2) \\ 
160 & B & IC348 & 03:43:50.5~+32:03:17 & 102 & 034385+32033 & 5 & 151 & 8.3(1) & 0.51(4) & 0.09(6) \\ 
161 & B & IC348 & 03:43:56.0~+32:00:45 & 103 & 034394+32008 & 5 & 153 & 9.5(1) & 0.70(4) & 0.15(6) \\ 
162 & B & IC348 & 03:43:57.3~+32:03:04 & 104 & 034395+32030 & 5 & 150 & 3.6(1) & 0.35(4) & 0.08(6) \\ 
163 & B & IC348 & 03:43:57.8~+32:04:06 & 105 & 034396+32040 & 15 & 57 & 3.62(4) & 0.36(2) & 0.06(2) \\ 
164 & B & IC348 & 03:44:01.7~+32:02:02 & 106 & 034402+32020 & 5 & 135 & 4.42(9) & 0.37(4) & 0.08(6) \\ 
165 & B & IC348 & 03:44:02.2~+32:02:32 & 107 & 034404+32025 & 10 & 55 & 6.74(4) & 0.36(2) & 0.01(2) \\ 
166 & B & IC348 & 03:44:02.3~+32:04:56 & 108 & \nodata & 15 & 54 & 0.77(4) & 0.15(2) & 0.00(2) \\ 
167 & D & IC348 & 03:44:04.6~+31:58:07 & \nodata & \nodata & 10 & 69 & 0.80(5) & 0.10(2) & 0.01(3) \\ 
168 & B & IC348 & 03:44:05.1~+32:00:28 & 109 & \nodata & 10 & 139 & 1.2(1) & 0.06(4) & 0.00(6) \\ 
169 & B & IC348 & 03:44:05.3~+32:02:05 & 110 & \nodata & 10 & 138 & 5.4(1) & 0.27(4) & 0.03(6) \\ 
170 & B & IC348 & 03:44:14.6~+31:57:59 & 111 & \nodata & 5 & 145 & 4.1(1) & 0.09(4) & 0.08(6) \\ 
171 & B & IC348 & 03:44:14.7~+32:09:11 & 112 & \nodata & 15 & 57 & 1.94(4) & 0.13(2) & -0.02(2) \\ 
172 & W & IC348 & 03:44:18.4~+32:06:36 & \nodata & \nodata & 5 & 179 & 0.0(1) & 0.00(5) & 0.00(7) \\ 
173 & B & IC348 & 03:44:22.6~+31:59:24 & 113 & \nodata & 5 & 144 & 6.4(1) & 0.23(4) & 0.09(6) \\ 
174 & B & IC348 & 03:44:22.6~+32:10:00 & 114 & \nodata & 15 & 71 & 1.43(5) & 0.16(2) & 0.01(3) \\ 
175 & D & IC348 & 03:44:30.0~+31:59:04 & \nodata & \nodata & 5 & 123 & 0.45(9) & 0.10(4) & -0.05(5) \\ 
176 & B & IC348 & 03:44:36.4~+31:58:40 & 115 & 034461+31587 & 10 & 65 & 2.89(5) & 0.13(2) & 0.04(2) \\ 
177 & W & IC348 & 03:44:37.7~+32:08:13 & \nodata & \nodata & 5 & 187 & -0.1(1) & -0.06(6) & -0.09(8) \\ 
178 & B & IC348 & 03:44:44.0~+32:01:24 & 116 & 034472+32015 & 10 & 136 & 1.62(9) & 0.24(4) & 0.05(5) \\ 
179 & W & IC348 & 03:44:46.2~+32:10:50 & \nodata & \nodata & 5 & 180 & 0.2(1) & -0.03(5) & -0.07(7) \\ 
180 & B & IC348 & 03:44:48.8~+32:00:29 & 117 & \nodata & 5 & 138 & 3.2(1) & 0.13(4) & -0.04(6) \\ 
181 & B & IC348 & 03:44:56.1~+32:00:32 & 118 & \nodata & 15 & 52 & 1.48(4) & 0.07(1) & 0.04(2) \\ 
182 & D & IC348 & 03:45:10.7~+32:00:38 & \nodata & \nodata & 10 & 69 & 0.36(5) & 0.01(2) & 0.01(3) \\ 
183 & B & IC348 & 03:45:15.9~+32:04:49 & 119 & \nodata & 5 & 144 & 3.5(1) & 0.14(4) & -0.02(6) \\ 
184 & B & \nodata & 03:45:48.0~+32:24:13 & 120 & N/A & 5 & 171 & 0.2(1) & -0.01(5) & 0.02(7) \\ 
185 & D & B5 & 03:46:43.0~+32:41:38 & \nodata & N/A & 5 & 123 & 0.33(9) & -0.01(3) & 0.01(5) \\ 
186 & W & B5 & 03:47:07.8~+32:42:01 & \nodata & N/A & 5 & 145 & 0.1(1) & 0.00(4) & 0.03(6) \\ 
187 & W & B5 & 03:47:22.0~+32:45:18 & \nodata & N/A & 5 & 138 & 0.2(1) & 0.04(4) & -0.04(6) \\ 
188 & B & B5 & 03:47:33.5~+32:50:55 & 121 & \nodata & 10 & 55 & 2.37(4) & 0.10(1) & 0.11(2) \\ 
189 & S & B5 & 03:47:38.6~+32:52:19 & \nodata & 034764+32523 & 5 & 154 & 6.0(1) & 0.21(4) & 0.20(6) \\ 
190 & L & B5 & 03:47:39.7~+32:53:57 & \nodata & \nodata & 10 & 53 & 1.88(4) & 0.09(2) & 0.36(2) \\ 
191 & D & B5 & 03:47:39.8~+32:53:34 & \nodata & \nodata & 15 & 77 & 2.69(5) & 0.08(2) & 0.29(3) \\ 
192 & S & B5 & 03:47:41.4~+32:51:48 & 122 & 034769+32517 & 5 & 171 & 6.8(1) & 0.44(4) & 0.06(6) \\ 
193 & W & B5 & 03:48:28.1~+32:50:15 & \nodata & N/A & 5 & 148 & -0.3(1) & 0.04(4) & -0.05(6) \\ 

\enddata

\tablecomments{(1) Running Source Number. (2) Origin of Source: B:
  BOLOCAM Core from \citet{bolocam-perseus}, S:SCUBA Core form
  \citet{perseus-scuba}, D: FIR Dust emission from
  \citet{schnee-mips}, L: Literature sources in \citet{jijina}. (3)
  Designation as defined in Figure \ref{findchart1}. (4) Position
  Observed. (5) Name of object in the \citet{bolocam-perseus} catalog.
  ``N/A'' is listed if the position is outside the boundaries of the
  BOLOCAM survey. (6) Name of object in the \citet{perseus-scuba}
  catalog.  ``N/A'' is listed if the position is outside the
  boundaries of the SCUBA survey.  (7) Total integration time on
  source. (8) Noise level in the NH$_3$(1,1) spectrum on the $T_{mb}$ scale.  (9)-(11) Integrated
  intensity of the observed lines on the $T_{mb}$ scale.}
\end{deluxetable}
\clearpage

%\end{landscape}

\begin{deluxetable}{ccccccccccccc}
\tablecaption{\label{proptable}Derived Physical Properties}
\tablewidth{0pt}
\tabletypesize{\footnotesize}
\tablehead{
%\begin{tabular}{cll}
\colhead{NH3SRC} & \colhead{$V_{LSR}$} & 
\colhead{$\sigma_v$} &
\colhead{$T_k$} &\colhead{$N_{\mathrm{NH3}}$\tablenotemark{a}} &
\colhead{$N_{\mathrm{CCS}}$\tablenotemark{a}} & \colhead{$\tau_{1}$} &
\colhead{$T_{x}$\tablenotemark{a}} & \colhead{$\eta_{f}$\tablenotemark{a}} & 
\colhead{$T_{\mathrm{CCS}}$\tablenotemark{a}} &
\colhead{$\sigma_{\mathrm{CCS}}$} &
\colhead{$V_{off}$} & \colhead{$\widetilde{\chi^2}$}
\\
\colhead{} & \colhead{(km s$^{-1}$)} & 
\colhead{(km s$^{-1}$)} &
\colhead{(K)} & \colhead{$10^{13}\mbox{ cm}^{-2}$} &
\colhead{$10^{12}\mbox{ cm}^{-2}$} & \colhead{} & \colhead{(K)} &
\colhead{} & \colhead{(K)} & \colhead{(km s$^{-1}$)} &
\colhead{(km s$^{-1}$)}\\
\colhead{(1)} & \colhead{(2)} & \colhead{(3)} & \colhead{(4)} &
\colhead{(5)} & \colhead{(6)} & \colhead{(7)} & \colhead{(8)} &
\colhead{(9)} & \colhead{(10)} & \colhead{(11)} & \colhead{(12)} & \colhead{(13)}
}
\startdata
 2 & 4.13(5) & $<0.20(6)$ &  $<26$  &  $>0.6$  & \nodata & \multicolumn{2}{c}{0.8(2)} & \nodata & \nodata & \nodata  & \nodata & 1.0 \\
 3 & 4.136(1) & 0.115(1) & 9.2(2) & 46.(2) & 4.0(4) & 6.6(2) & 6.47(7) & 0.58 & 0.51(4) & 0.15(1)  & 0.02(1) & 0.9 \\
 4 & 3.949(2) & 0.155(2) & 9.2(2) & 53.(2) & 2.3(4) & 6.1(2) & 6.14(6) & 0.52 & 0.29(4) & 0.15(2)  & 0.06(2) & 1.1 \\
 5 & 3.982(2) & 0.135(2) & 10.0(2) & 26.(2) & 5.2(8) & 3.8(2) & 6.5(2) & 0.52 & 0.34(3) & 0.26(3)  & -0.14(3) & 1.4 \\
 6 & 4.030(3) & 0.134(3) & 10.3(4) & 12.(2) & 8.1(6) & 1.8(3) & 6.4(5) & 0.48 & 0.67(3) & 0.20(1)  & 0.09(1) & 1.0 \\
 7 & 4.135(1) & 0.168(1) & 12.06(8) & 54.(1) & 5.1(7) & 7.4(1) & 7.45(4) & 0.50 & 0.30(3) & 0.23(3)  & 0.05(3) & 1.2 \\
 8 & 4.090(1) & 0.155(1) & 11.21(9) & 84.(2) & 7.(1) & 10.9(2) & 7.61(4) & 0.57 & 0.53(5) & 0.19(2)  & 0.09(2) & 1.0 \\
 9 & 4.563(2) & 0.124(2) & 9.1(2) & 39.(3) & 3.7(5) & 5.5(2) & 6.04(9) & 0.52 & 0.44(4) & 0.16(2)  & -0.05(2) & 1.0 \\
 10 & 4.295(9) & 0.29(1) & 11.1(4) & 28.(4) & \nodata & 3.3(4) & 4.3(1) & 0.18 & \nodata & \nodata  & \nodata & 1.0 \\
 11 & 4.236(8) & 0.143(8) &  $<13$  &  $>5$  &  $<12$  & 1.2(7) & 5.(1) & 0.27 & 0.86(3) & 0.163(8)  & -0.01(1) & 0.9 \\
 12\tablenotemark{b} & 4.523(2) & $<0.373(2)$ & 12.61(5) & 91.(2) & 7.2(8) & 5.19(7) & 8.55(4) & 0.59 & 0.22(2) & 0.42(4)  & -0.05(4) & 2.7 \\
 13 & 4.06(2) & $<0.24(2)$ &  $<15$  &  $>7$  &  $<4.3$  & 1.5(7) & 3.5(3) & 0.12 & 0.24(2) & 0.19(2)  & 0.06(3) & 0.9 \\
 14 & 5.074(3) & 0.392(3) & 14.05(9) & 55.(2) & 10.(1) & 3.60(9) & 7.83(7) & 0.45 & 0.46(3) & 0.24(2)  & -0.42(2) & 1.6 \\
 15 & 4.648(3) & 0.194(3) & 11.2(2) & 31.(2) & 4.9(9) & 3.7(2) & 6.6(1) & 0.46 & 0.25(3) & 0.29(4)  & -0.06(4) & 1.0 \\
 16 & 4.614(7) & 0.141(6) &  $<13$  &  $>1.3$  &  $<12$  & \multicolumn{2}{c}{1.85(8)} & \nodata & 1.07(3) & 0.132(4)  & 0.026(8) & 1.0 \\
 17\tablenotemark{b} & 4.5058(7) & 0.1456(6) & 9.13(5) & 133.(2) & 4.8(4) & 14.7(2) & 6.71(2) & 0.62 & 0.30(2) & 0.32(2)  & -0.22(2) & 4.4 \\
 18 & 4.539(1) & 0.153(1) & 9.0(1) & 119.(4) & 5.0(7) & 12.7(2) & 6.44(3) & 0.59 & 0.34(3) & 0.29(3)  & -0.10(3) & 3.3 \\
 19 & 4.345(7) & 0.256(9) & 10.0(6) & 13.(4) & 4.7(5) & 1.4(3) & 4.7(4) & 0.27 & 0.37(2) & 0.22(2)  & -0.03(2) & 0.9 \\
 20 & 4.17(1) & $<0.28(2)$ &  $<14$  &  $>1.3$  &  $<6.0$  & \multicolumn{2}{c}{1.02(5)} & \nodata & 0.43(2) & 0.16(1)  & 0.06(2) & 0.9 \\
 21 & 3.559(9) & $<0.21(1)$ &  $<13.6$  &  $>1.5$  &  $<2.0$  & \multicolumn{2}{c}{1.46(7)} & \nodata & 0.12(2) & 0.19(4)  & -0.02(4) & 0.9 \\
 22 & 5.150(2) & 0.146(2) & 11.7(2) & 32.(1) & 6.4(7) & 6.3(2) & 5.63(6) & 0.32 & 0.57(4) & 0.16(1)  & 0.04(1) & 1.0 \\
 24 & 4.591(3) & 0.149(3) & 10.4(4) & 13.(2) & 5.5(6) & 2.2(3) & 5.1(3) & 0.32 & 0.53(3) & 0.17(1)  & 0.03(1) & 0.8 \\
 25 & 0.90(4) & $<0.20(5)$ & \nodata & \nodata & \nodata & \multicolumn{2}{c}{1.1(2)} & \nodata & \nodata & \nodata  & \nodata & 1.1 \\
 26 & 5.32(3) & $<0.32(3)$ &  $<18$  &  $>7$  & \nodata & 1.(1) & 3.7(6) & 0.10 & \nodata & \nodata  & \nodata & 1.0 \\
 27 & 5.26(3) & $<0.22(4)$ &  $<20$  &  $>1.0$  &  $<4$  & \multicolumn{2}{c}{1.2(2)} & \nodata & 0.29(8) & 0.11(4)  & 0.05(5) & 1.0 \\
 28 & 5.62(1) & $<0.11(1)$ &  $<18$  &  $>3$  &  $<9$  & 2.(2) & 5.(2) & 0.29 & 0.9(1) & 0.08(1)  & 0.09(2) & 1.0 \\
 29 & 5.083(2) & 0.125(1) & 10.7(2) & 27.(1) & 6.6(4) & 4.8(2) & 6.4(1) & 0.46 & 1.00(4) & 0.105(5)  & -0.352(5) & 1.7 \\
 30 & 5.22(1) & $<0.18(1)$ &  $<13$  &  $>9$  &  $<3.7$  & 1.9(6) & 4.6(5) & 0.25 & 0.41(6) & 0.11(2)  & -0.10(2) & 0.9 \\
31.1 & 4.6(2) & 0.17(1) & 11.7(1) & 23.(2) & 2.7(3) & 3.749(2) & \nodata & 0.43 & 0.120(5) & 0.32(3)  & 0.10(6) & 4.1 \\
31.2 & 6.01(5) & 0.13(2) & 10.4(1) & 10.(1) & 1.8(3) & 1.739(2) & \nodata & 0.24 & 0.175(6) & 0.17(3)  & 0.02(2) & 4.1 \\
 32 & 4.759(2) & 0.275(2) & 13.05(7) & 48.(1) & 5.7(8) & 4.40(8) & 7.46(5) & 0.46 & 0.22(2) & 0.31(3)  & 0.09(3) & 2.3 \\
 33 & 4.928(3) & 0.186(3) & 10.5(2) & 79.(4) & 7.(3) & 7.8(3) & 7.65(9) & 0.63 & 0.27(6) & 0.4(1)  & 0.2(1) & 1.0 \\
 34\tablenotemark{b} & 4.9691(9) & 0.1690(8) & 11.37(6) & 54.(1) & 5.2(6) & 6.08(9) & 8.32(4) & 0.65 & 0.37(2) & 0.21(1)  & 0.02(1) & 3.6 \\
 35 & 4.949(2) & 0.211(2) & 11.9(1) & 32.(1) & 3.5(8) & 3.7(1) & 6.70(7) & 0.43 & 0.10(2) & 0.47(8)  & 0.21(8) & 1.1 \\
 36 & 4.706(1) & 0.083(1) & 9.1(2) & 48.(3) & 2.6(9) & 9.7(4) & 6.33(7) & 0.57 & 0.20(4) & 0.26(6)  & 0.02(6) & 1.0 \\
 37 & 4.861(3) & 0.115(3) & 10.3(4) & 27.(4) & 4.(1) & 5.0(5) & 6.2(2) & 0.45 & 0.40(9) & 0.17(4)  & 0.01(4) & 0.9 \\
 38 & 4.940(2) & 0.126(2) & 9.4(3) & 44.(3) & 2.8(7) & 6.6(3) & 5.99(9) & 0.49 & 0.28(5) & 0.18(3)  & 0.22(3) & 0.9 \\
 40 & 7.193(1) & 0.151(1) & 10.8(1) & 41.(1) & 1.6(4) & 6.9(2) & 5.62(4) & 0.36 & 0.14(2) & 0.19(4)  & 0.06(4) & 1.0 \\
 41 & 6.655(2) & 0.148(1) & 10.6(1) & 39.(1) & 2.7(7) & 5.4(2) & 6.83(7) & 0.52 & 0.13(3) & 0.33(7)  & -0.06(7) & 1.4 \\
 42 & 5.471(4) & 0.136(4) & 10.7(3) & 18.(2) & 17.(1) & 3.8(4) & 4.9(2) & 0.27 & 1.85(3) & 0.150(3)  & 0.125(5) & 1.3 \\
 43 & 6.823(2) & 0.110(2) & 10.5(2) & 25.(2) & 1.2(6) & 6.0(3) & 5.21(8) & 0.32 & 0.19(5) & 0.10(3)  & 0.06(3) & 1.0 \\
 44 & 7.338(2) & 0.216(2) & 12.4(1) & 27.(1) & 1.5(5) & 3.6(1) & 6.13(7) & 0.35 & 0.11(3) & 0.18(5)  & 0.06(5) & 1.3 \\
 46 & 7.030(1) & 0.152(1) & 10.54(9) & 56.(2) & 4.1(7) & 7.0(1) & 7.35(5) & 0.59 & 0.33(3) & 0.20(2)  & 0.07(2) & 1.3 \\
 47 & 8.1840(7) & 0.1886(6) & 11.69(4) & 73.2(9) & 4.9(4) & 8.15(7) & 7.82(2) & 0.57 & 0.29(2) & 0.23(2)  & 0.15(2) & 2.6 \\
 48 & 7.985(3) & 0.192(2) & 11.7(1) & 91.(4) & \nodata & 8.8(3) & 9.01(9) & 0.70 & \nodata & \nodata  & \nodata & 1.0 \\
 49 & 5.311(9) & $<0.21(1)$ &  $<13$  &  $>1.8$  &  $<12$  & \multicolumn{2}{c}{1.66(7)} & \nodata & 0.99(3) & 0.159(5)  & 0.05(1) & 1.1 \\
 50 & 7.213(1) & 0.158(1) & 10.5(1) & 60.(2) & 3.4(8) & 7.8(2) & 6.69(4) & 0.51 & 0.21(3) & 0.26(4)  & 0.12(4) & 1.1 \\
 51 & 8.113(2) & 0.194(2) & 10.8(1) & 49.(2) & \nodata & 4.9(1) & 7.63(8) & 0.61 & \nodata & \nodata  & \nodata & 1.3 \\
 52 & 8.056(1) & 0.124(1) & 11.3(1) & 40.(1) & \nodata & 7.1(2) & 7.18(7) & 0.52 & \nodata & \nodata  & \nodata & 1.1 \\
 53 & 5.84(2) & $<0.39(3)$ &  $<17$  &  $>0.94$  &  $<4.3$  & \multicolumn{2}{c}{0.59(3)} & \nodata & 0.38(3) & 0.10(1)  & -0.14(3) & 1.0 \\
 54 & 7.91(4) & $<0.16(4)$ & \nodata & \nodata & \nodata & \multicolumn{2}{c}{0.5(1)} & \nodata & \nodata & \nodata  & \nodata & 1.0 \\
 55 & 5.39(3) & $<0.22(4)$ &  $<19$  &  $>11$  & \nodata & 3.(2) & 3.2(2) & 0.08 & \nodata & \nodata  & \nodata & 1.0 \\
 56 & 7.623(3) & 0.192(3) & 13.8(2) & 20.(2) & \nodata & 2.5(2) & 8.4(3) & 0.51 & \nodata & \nodata  & \nodata & 1.0 \\
 57 & 7.85(1) & 0.35(1) & 16.4(7) & 3.3(2) & \nodata & \multicolumn{2}{c}{2.26(8)} & \nodata & \nodata & \nodata  & \nodata & 0.9 \\
 58 & 7.476(3) & 0.372(3) & 16.5(1) & 34.(2) & 6.(3) & 2.5(1) & 8.4(1) & 0.41 & 0.14(3) & 0.4(1)  & 0.4(1) & 1.9 \\
 59 & 7.778(2) & 0.172(2) & 14.5(1) & 28.(1) & \nodata & 4.2(1) & 8.1(1) & 0.45 & \nodata & \nodata  & \nodata & 1.0 \\
 60 & 5.73(3) & $<0.18(3)$ &  $<19$  &  $>7$  &  $<7$  & 2.(2) & 3.3(3) & 0.09 & 0.36(6) & 0.15(3)  & 0.04(4) & 0.9 \\
 61 & 8.1(1) & $<0.7(1)$ &  $<29$  &  $>0.7$  & \nodata & \multicolumn{2}{c}{0.29(4)} & \nodata & \nodata & \nodata  & \nodata & 1.0 \\
 62 & 5.82(5) & $<0.26(6)$ & \nodata & \nodata & \nodata & \multicolumn{2}{c}{0.7(1)} & \nodata & \nodata & \nodata  & \nodata & 0.9 \\
 63 & 7.8(1) & $<0.7(1)$ &  $<26$  &  $>0.9$  & \nodata & \multicolumn{2}{c}{0.37(6)} & \nodata & \nodata & \nodata  & \nodata & 1.0 \\
 64 & 7.794(3) & 0.307(4) & 14.4(2) & 27.(2) & \nodata & 2.2(1) & 8.6(2) & 0.50 & \nodata & \nodata  & \nodata & 0.9 \\
 65 & 7.147(2) & 0.208(2) & 12.5(1) & 37.(1) & 1.9(7) & 4.5(1) & 6.98(8) & 0.43 & 0.15(4) & 0.17(5)  & 0.13(5) & 1.2 \\
 66 & 8.006(2) & 0.269(3) & 16.4(1) & 33.(1) & \nodata & 2.8(1) & 10.0(2) & 0.53 & \nodata & \nodata  & \nodata & 0.9 \\
 67 & 8.436(3) & 0.296(4) & 16.3(2) & 41.(2) & \nodata & 3.6(2) & 8.7(2) & 0.44 & \nodata & \nodata  & \nodata & 0.9 \\
 68\tablenotemark{b} & 7.433(5) & $<0.578(5)$ & 16.4(1) & 52.(3) & \nodata & 2.3(1) & 9.0(2) & 0.46 & \nodata & \nodata  & \nodata & 1.6 \\
 69 & 8.283(3) & 0.247(3) & 13.6(2) & 25.(1) & \nodata & 2.7(1) & 7.3(2) & 0.42 & \nodata & \nodata  & \nodata & 0.9 \\
 70 & 8.028(4) & 0.357(4) & 14.8(2) & 46.(2) & \nodata & 3.2(1) & 8.5(2) & 0.48 & \nodata & \nodata  & \nodata & 1.3 \\
 71 & 7.81(5) & $<0.72(5)$ & 23.(2) & 1.16(9) & \nodata & \multicolumn{2}{c}{0.44(3)} & \nodata & \nodata & \nodata  & \nodata & 0.9 \\
 72 & 8.483(3) & 0.208(3) & 12.6(2) & 29.(2) & 2.(1) & 3.3(2) & 7.7(1) & 0.50 & 0.11(4) & 0.21(9)  & -0.06(9) & 1.1 \\
 73\tablenotemark{b} & 7.760(2) & $<0.462(2)$ & 12.32(6) & 89.(1) & 5.(1) & 4.09(6) & 8.31(5) & 0.58 & 0.13(2) & 0.50(9)  & 0.02(9) & 3.3 \\
 75 & 7.335(5) & 0.595(5) & 15.0(1) & 43.(2) & 5.(1) & 2.2(1) & 7.1(1) & 0.36 & 0.15(3) & 0.32(6)  & 0.57(6) & 1.1 \\
 76 & 7.7(1) & $<0.7(1)$ &  $<35$  &  $>1.5$  & \nodata & \multicolumn{2}{c}{0.58(9)} & \nodata & \nodata & \nodata  & \nodata & 1.0 \\
 77 & 8.574(2) & 0.222(2) & 14.3(1) & 34.(1) & \nodata & 3.5(1) & 9.1(1) & 0.55 & \nodata & \nodata  & \nodata & 1.4 \\
 78 & 7.12(3) & 0.56(3) & 13.5(7) & 40(1) & \nodata & 2.1(5) & 5.8(5) & 0.28 & \nodata & \nodata  & \nodata & 0.8 \\
 79 & 8.26(1) & 0.42(1) & 15.5(5) & 3.4(1) & \nodata & \multicolumn{2}{c}{1.91(4)} & \nodata & \nodata & \nodata  & \nodata & 0.8 \\
80.1 & 7.6(5) & $<0.28(2)$ & 11.6(3) & 23.(2) & \nodata & 2.26(1) & \nodata & 0.30 & \nodata & \nodata  & \nodata & 1.6 \\
80.2 & 8.1(5) & $<0.12(2)$ & 10.1(2) & 35.(7) & \nodata & 6.33(2) & \nodata & 0.22 & \nodata & \nodata  & \nodata & 1.6 \\
 81 & 7.487(2) & 0.129(2) & 11.8(2) & 21.(1) & \nodata & 4.2(2) & 6.5(1) & 0.41 & \nodata & \nodata  & \nodata & 0.9 \\
82.1 & 7.0(2) & 0.15(1) & 11.9(2) & 20.(2) & \nodata & 4.241(3) & \nodata & 0.28 & \nodata & \nodata  & \nodata & 2.3 \\
82.2 & 7.5(2) & 0.1(1) & 13.4(3) & 12.(9) & \nodata & 3.710(3) & \nodata & 0.15 & \nodata & \nodata  & \nodata & 2.3 \\
83.1 & 7.5(2) & $<0.20(7)$ & 11.2(3) & 11.(4) & 1.9(2) & 1.339(6) & \nodata & 0.47 & 0.085(2) & 0.34(2)  & 0.10(2) & 3.6 \\
83.2 & 8.1(1) & $<0.10(2)$ & 9.9(3) & 31.(8) & \nodata & 6.114(7) & \nodata & 0.36 & \nodata & \nodata  & \nodata & 3.6 \\
 84 & 7.478(4) & 0.146(4) & 13.3(4) & 15.(2) & \nodata & 4.0(4) & 4.8(2) & 0.19 & \nodata & \nodata  & \nodata & 0.9 \\
 85 & 8.60(7) & $<0.55(7)$ &  $<22$  &  $>0.8$  & \nodata & \multicolumn{2}{c}{0.38(4)} & \nodata & \nodata & \nodata  & \nodata & 1.0 \\
 86 & 7.202(3) & 0.136(3) & 9.8(4) & 22.(3) & 0.7(6) & 3.6(3) & 5.6(2) & 0.40 & 0.16(6) & 0.07(3)  & -0.02(3) & 0.8 \\
 87 & 7.495(1) & 0.126(1) & 10.4(1) & 44.(2) & 1.3(6) & 6.6(2) & 7.28(7) & 0.59 & 0.20(5) & 0.11(3)  & 0.09(3) & 0.9 \\
88.1 & 7.4(2) & 0.13(1) & 10.3(2) & 40.(5) & \nodata & 6.444(3) & \nodata & 0.46 & \nodata & \nodata  & \nodata & 2.3 \\
88.2 & 7.77(7) & 0.09(5) & 10.5(2) & 50(3) & 0.2(2) & 12.578(3) & \nodata & 0.17 & 0.113(4) & 0.03(3)  & -0.1(1) & 2.3 \\
 89 & 8.179(1) & 0.125(1) & 10.5(1) & 39.(2) & 2.1(9) & 5.5(2) & 7.9(1) & 0.66 & 0.18(4) & 0.19(5)  & 0.03(5) & 1.2 \\
 90 & 7.88(7) & $<0.34(8)$ &  $<27$  &  $>0.7$  & \nodata & \multicolumn{2}{c}{0.5(1)} & \nodata & \nodata & \nodata  & \nodata & 0.9 \\
 91 & 5.882(1) & 0.142(1) & 11.0(1) & 49.(2) & \nodata & 7.5(2) & 6.85(6) & 0.50 & \nodata & \nodata  & \nodata & 1.1 \\
 93 & 6.022(1) & 0.098(1) & 10.5(2) & 14.(1) & 6.4(4) & 3.3(2) & 6.1(2) & 0.43 & 0.84(3) & 0.123(5)  & 0.019(5) & 1.0 \\
 95 & 6.0635(9) & 0.1452(8) & 10.01(6) & 106.(2) & 4.1(5) & 13.8(2) & 6.75(2) & 0.55 & 0.47(4) & 0.15(1)  & -0.07(1) & 1.4 \\
 96 & 7.838(3) & 0.152(3) & 10.5(3) & 19.(2) & 8.2(7) & 3.1(3) & 5.4(2) & 0.34 & 0.95(4) & 0.141(7)  & 0.013(8) & 1.0 \\
 97 & 7.724(2) & 0.102(2) & 9.8(3) & 20.(2) & 2.8(4) & 4.2(3) & 5.5(2) & 0.40 & 0.48(4) & 0.10(1)  & 0.03(1) & 0.9 \\
 98 & 7.02(3) & $<0.34(4)$ &  $<20$  &  $>1.7$  & \nodata & \multicolumn{2}{c}{1.3(1)} & \nodata & \nodata & \nodata  & \nodata & 1.0 \\
 99 & 6.992(1) & 0.182(1) & 11.44(7) & 49.6(8) & 1.5(5) & 5.97(9) & 7.26(4) & 0.52 & 0.09(2) & 0.24(6)  & -0.07(6) & 1.6 \\
 101 & 6.78(5) & $<0.22(5)$ &  $<23$  &  $>6$  & \nodata & 2.(2) & 3.4(6) & 0.07 & \nodata & \nodata  & \nodata & 0.9 \\
 102 & 6.07(6) & $<0.32(8)$ &  $<26$  &  $>0.8$  &  $<10$  & \multicolumn{2}{c}{0.7(1)} & \nodata & 0.5(1) & 0.10(3)  & -0.15(7) & 1.0 \\
 103 & 6.881(2) & 0.243(2) & 11.70(9) & 62.(1) & 6.(1) & 5.6(1) & 7.53(5) & 0.53 & 0.40(4) & 0.22(3)  & 0.19(3) & 1.3 \\
 104 & 6.420(1) & 0.153(1) & 10.48(8) & 52.(1) & 16.6(5) & 7.7(1) & 6.07(3) & 0.43 & 1.23(2) & 0.220(4)  & 0.041(4) & 1.5 \\
 105 & 6.646(2) & 0.107(1) & 9.6(2) & 36.(3) & 2.7(6) & 7.7(3) & 5.09(6) & 0.34 & 0.22(3) & 0.22(3)  & 0.17(3) & 1.1 \\
 107\tablenotemark{b} & 6.40(2) & $<0.53(2)$ &  $<13$  &  $>2.7$  &  $<6$  & \multicolumn{2}{c}{1.05(3)} & \nodata & 0.15(2) & 0.49(6)  & 0.05(6) & 0.9 \\
 108 & 6.804(1) & 0.1283(9) & 10.16(9) & 72.(2) & 4.7(5) & 11.1(2) & 6.63(3) & 0.52 & 0.51(4) & 0.15(1)  & -0.02(1) & 1.3 \\
 109 & 6.579(6) & 0.230(7) & 11.2(4) & 11.(2) & 4.8(5) & 1.4(3) & 5.0(4) & 0.26 & 0.46(3) & 0.16(1)  & 0.17(1) & 0.9 \\
 110 & 6.82(3) & $<0.30(3)$ &  $<18$  &  $>1.1$  &  $<3$  & \multicolumn{2}{c}{0.94(8)} & \nodata & 0.25(5) & 0.11(3)  & 0.18(4) & 0.9 \\
 111 & 6.673(1) & 0.139(1) & 10.1(1) & 43.(1) & 5.2(3) & 6.2(1) & 6.41(5) & 0.50 & 0.54(2) & 0.161(9)  & 0.055(9) & 1.0 \\
 112 & 6.583(7) & 0.257(8) & 11.2(5) & 11.(3) & 3.8(4) & 1.2(3) & 5.2(5) & 0.29 & 0.40(3) & 0.14(1)  & 0.20(1) & 0.9 \\
 113 & 6.570(1) & 0.153(1) & 10.0(1) & 61.(2) & 8.7(4) & 7.6(2) & 6.68(5) & 0.54 & 0.83(3) & 0.182(8)  & 0.100(8) & 1.1 \\
 114 & 6.609(1) & 0.1894(9) & 9.86(6) & 106.(2) & 8.7(5) & 9.9(1) & 7.04(2) & 0.60 & 0.83(3) & 0.184(7)  & 0.107(7) & 1.8 \\
 115 & 6.574(6) & 0.164(6) & 10.3(5) & 33.(6) & 3.(1) & 4.6(5) & 5.8(2) & 0.40 & 0.4(1) & 0.13(4)  & 0.01(4) & 0.9 \\
 116 & 6.80(4) & $<0.57(5)$ &  $<16$  &  $>17$  &  $<4.2$  & 1.9(9) & 3.2(2) & 0.05 & 0.49(6) & 0.08(1)  & 0.33(5) & 0.9 \\
 117 & 6.86(2) & $<0.32(3)$ &  $<17$  &  $>1.3$  &  $<4$  & \multicolumn{2}{c}{1.02(7)} & \nodata & 0.09(3) & 0.4(1)  & -0.1(1) & 0.9 \\
 118 & 6.826(1) & 0.1304(9) & 9.39(9) & 116.(3) & 2.1(6) & 15.9(3) & 6.29(3) & 0.53 & 0.23(4) & 0.17(4)  & 0.08(4) & 1.4 \\
 119 & 6.409(1) & 0.260(1) & 11.52(4) & 100.(1) & 12.1(7) & 8.10(7) & 7.66(2) & 0.56 & 0.41(2) & 0.43(2)  & 0.31(2) & 2.4 \\
 121\tablenotemark{b} & 6.251(1) & $<0.331(1)$ & 12.43(4) & 81.(1) & 12.(1) & 5.36(6) & 8.17(3) & 0.56 & 0.34(2) & 0.46(2)  & 0.24(2) & 7.1 \\
 122 & 6.86(4) & $<0.33(4)$ &  $<22$  &  $>0.8$  & \nodata & \multicolumn{2}{c}{0.65(7)} & \nodata & \nodata & \nodata  & \nodata & 1.0 \\
 123 & 6.604(1) & 0.334(1) & 11.71(5) & 113.(2) & 11.7(8) & 7.21(8) & 7.72(3) & 0.56 & 0.45(2) & 0.36(2)  & 0.02(2) & 2.9 \\
 124 & 6.786(1) & 0.121(1) & 9.7(1) & 61.(2) & 2.0(5) & 10.4(3) & 5.88(4) & 0.45 & 0.28(4) & 0.13(2)  & 0.06(2) & 1.2 \\
 125 & 6.48(1) & $<0.40(1)$ & 12.0(4) & 16.(4) & 4.5(8) & 1.5(3) & 4.3(2) & 0.17 & 0.30(3) & 0.21(2)  & -0.08(2) & 1.0 \\
 126 & 6.790(4) & 0.254(5) & 11.9(3) & 19.(2) & 7.7(5) & 1.9(2) & 6.4(3) & 0.40 & 0.77(4) & 0.139(8)  & 0.034(9) & 1.1 \\
 127\tablenotemark{b} & 6.304(6) & $<0.260(7)$ & 10.9(3) & 22.(3) & \nodata & 2.3(3) & 5.2(2) & 0.30 & \nodata & \nodata  & \nodata & 1.7 \\
 128 & 7.678(2) & 0.079(2) &  $<10.7$  &  $>13$  &  $<1.1$  & 5.2(5) & 4.2(1) & 0.22 & 0.27(4) & 0.07(1)  & 0.06(1) & 0.9 \\
 129 & 7.59(3) & $<0.32(4)$ &  $<20$  &  $>1.4$  & \nodata & \multicolumn{2}{c}{1.2(1)} & \nodata & \nodata & \nodata  & \nodata & 1.0 \\
 132 & 7.06(1) & $<0.17(1)$ &  $<15$  &  $>5$  &  $<3$  & 1.6(9) & 3.8(5) & 0.13 & 0.24(5) & 0.12(3)  & 0.05(3) & 0.9 \\
 139 & 5.29(4) & 0.43(4) & 24.(2) & 0.64(8) & \nodata & \multicolumn{2}{c}{0.40(4)} & \nodata & \nodata & \nodata  & \nodata & 0.9 \\
 141 & 8.19(2) & $<0.16(2)$ &  $<16$  &  $>6$  &  $<6$  & 2.(1) & 5.(1) & 0.29 & 0.4(1) & 0.14(4)  & 0.02(4) & 0.9 \\
 142 & 8.475(3) & 0.113(3) & 12.4(4) & 10.(2) & 3.3(8) & 3.0(5) & 5.1(3) & 0.24 & 0.35(5) & 0.12(2)  & 0.07(2) & 1.0 \\
 144 & 8.24(2) & 0.21(2) & 20.(2) & 0.57(9) & \nodata & \multicolumn{2}{c}{0.73(7)} & \nodata & \nodata & \nodata  & \nodata & 0.9 \\
 145 & 9.431(3) & 0.094(2) & 9.5(4) & 43.(6) & 2.(1) & 8.6(7) & 6.1(2) & 0.50 & 0.4(1) & 0.10(3)  & 0.07(3) & 0.9 \\
 146 & 8.18(1) & $<0.18(1)$ & 17.(1) & 5.(3) & \nodata & 1.4(9) & 3.7(5) & 0.07 & \nodata & \nodata  & \nodata & 1.0 \\
 147 & 9.432(1) & 0.0980(9) & 9.6(1) & 52.(2) & 1.5(5) & 9.6(3) & 6.54(6) & 0.55 & 0.28(5) & 0.10(2)  & 0.02(2) & 0.9 \\
 148 & 9.43(2) & $<0.15(2)$ &  $<18$  &  $>10$  & \nodata & 4.(2) & 4.1(5) & 0.14 & \nodata & \nodata  & \nodata & 0.9 \\
 150 & 8.54(5) & $<0.47(6)$ &  $<20$  &  $>7$  & \nodata & 1.(1) & 3.1(4) & 0.04 & \nodata & \nodata  & \nodata & 1.0 \\
 151 & 8.64(2) & $<0.28(3)$ &  $<17$  &  $>0.9$  &  $<3$  & \multicolumn{2}{c}{0.84(6)} & \nodata & 0.20(4) & 0.14(3)  & 0.06(4) & 0.9 \\
 152 & 8.82(1) & 0.14(1) & 14.(2) & 6.(4) & 4.(2) & 2.(1) & 5.(1) & 0.17 & 0.5(1) & 0.10(3)  & 0.06(3) & 0.9 \\
 153 & 8.64(5) & $<0.20(5)$ &  $<23$  &  $>9$  & \nodata & 3.(3) & 3.3(3) & 0.05 & \nodata & \nodata  & \nodata & 0.9 \\
 154 & 8.71(5) & $<0.39(6)$ & 17.(3) & 1.4(3) & \nodata & \multicolumn{2}{c}{0.9(1)} & \nodata & \nodata & \nodata  & \nodata & 1.0 \\
 156 & 8.55(1) & 0.35(2) & 13.5(6) & 19.(5) & \nodata & 1.7(4) & 6.0(6) & 0.31 & \nodata & \nodata  & \nodata & 0.9 \\
157.1 & 8.3(2) & 0.16(3) & 12.9(6) & 12.(2) & \nodata & 2.150(9) & \nodata & 0.15 & \nodata & \nodata  & \nodata & 1.9 \\
157.2 & 8.7(6) & 0.14(3) & 12.3(5) & 15.(3) & 1.4(5) & 2.95(1) & \nodata & 0.33 & 0.156(3) & 0.12(3)  & 0.07(3) & 1.9 \\
 158 & 9.011(4) & 0.134(4) & 14.8(4) & 2.3(1) & 2.3(5) & \multicolumn{2}{c}{3.8(1)} & \nodata & 0.41(6) & 0.058(9)  & 0.06(1) & 1.3 \\
 159 & 8.709(2) & 0.180(2) & 11.7(1) & 28.(1) & 1.5(6) & 3.5(1) & 7.2(1) & 0.50 & 0.16(4) & 0.14(4)  & 0.12(4) & 1.1 \\
 160 & 8.637(4) & 0.195(4) & 11.7(3) & 36.(3) & \nodata & 3.5(3) & 8.6(3) & 0.65 & \nodata & \nodata  & \nodata & 0.9 \\
 161 & 8.984(5) & 0.224(5) & 12.9(3) & 40.(3) & \nodata & 4.3(3) & 7.6(2) & 0.48 & \nodata & \nodata  & \nodata & 0.9 \\
 162 & 8.742(9) & 0.22(1) & 14.2(6) & 13.(4) & \nodata & 1.7(5) & 7.(1) & 0.39 & \nodata & \nodata  & \nodata & 0.9 \\
 163 & 8.269(3) & 0.179(3) & 12.9(2) & 20.(1) & 1.9(6) & 3.8(2) & 5.3(1) & 0.25 & 0.18(4) & 0.14(3)  & 0.02(3) & 1.0 \\
 164 & 9.003(9) & 0.27(1) & 13.0(5) & 17.(4) & \nodata & 1.5(4) & 7.6(9) & 0.48 & \nodata & \nodata  & \nodata & 0.8 \\
 165 & 8.471(2) & 0.184(2) & 11.1(1) & 41.(2) & \nodata & 5.5(2) & 6.21(6) & 0.41 & \nodata & \nodata  & \nodata & 3.2 \\
 166 & 8.28(3) & 0.38(4) & 26.(2) & 0.70(9) & \nodata & \multicolumn{2}{c}{0.51(4)} & \nodata & \nodata & \nodata  & \nodata & 1.0 \\
 167 & 8.28(4) & $<0.54(5)$ &  $<21$  &  $>1.1$  & \nodata & \multicolumn{2}{c}{0.56(4)} & \nodata & \nodata & \nodata  & \nodata & 1.0 \\
 168 & 8.14(2) & 0.22(3) & 11.(2) & 7.(7) & \nodata & 1.(1.) & 5.(3) & 0.45 & \nodata & \nodata  & \nodata & 1.0 \\
 169 & 8.473(4) & 0.154(4) & 10.9(4) & 32.(4) & \nodata & 4.6(4) & 6.5(2) & 0.46 & \nodata & \nodata  & \nodata & 0.9 \\
 170 & 8.983(5) & 0.137(5) & 10.6(6) & 21.(5) & \nodata & 3.1(5) & 6.9(5) & 0.53 & \nodata & \nodata  & \nodata & 0.9 \\
 171 & 7.920(5) & 0.160(5) & 12.9(4) & 8.(2) & \nodata & 1.6(4) & 5.7(5) & 0.29 & \nodata & \nodata  & \nodata & 0.9 \\
 173 & 9.134(4) & 0.153(4) & 10.5(4) & 33.(4) & \nodata & 3.9(4) & 7.8(3) & 0.65 & \nodata & \nodata  & \nodata & 0.9 \\
 174 & 7.59(2) & $<0.37(2)$ & 14.7(9) & 2.6(2) & \nodata & \multicolumn{2}{c}{1.56(7)} & \nodata & \nodata & \nodata  & \nodata & 0.9 \\
 176 & 9.968(4) & 0.166(4) & 10.8(3) & 17.(3) & \nodata & 2.5(3) & 5.7(3) & 0.37 & \nodata & \nodata  & \nodata & 1.1 \\
 178 & 9.89(2) & 0.34(3) & 16.(1) & 2.7(3) & \nodata & \multicolumn{2}{c}{1.9(1)} & \nodata & \nodata & \nodata  & \nodata & 1.0 \\
 180 & 8.954(5) & 0.121(4) & 10.8(6) & 21.(4) & \nodata & 4.3(6) & 5.8(3) & 0.37 & \nodata & \nodata  & \nodata & 0.9 \\
 181 & 9.096(4) & 0.138(4) & 10.9(5) & 8.(3) & 0.8(4) & 1.3(4) & 6.2(9) & 0.43 & 0.08(3) & 0.16(8)  & 0.05(8) & 1.0 \\
 182 & 9.14(3) & $<0.19(4)$ &  $<22$  &  $>0.5$  & \nodata & \multicolumn{2}{c}{0.7(1)} & \nodata & \nodata & \nodata  & \nodata & 1.0 \\
 183 & 9.992(4) & 0.109(3) & 10.7(5) & 23.(4) & \nodata & 5.0(6) & 6.0(3) & 0.41 & \nodata & \nodata  & \nodata & 0.9 \\
 188 & 10.151(2) & 0.107(2) & 10.0(3) & 15.(2) & \nodata & 2.6(3) & 6.6(3) & 0.53 & \nodata & \nodata  & \nodata & 1.0 \\
 189 & 10.370(6) & 0.182(5) & 10.1(4) & 44.(5) & 5.(2) & 5.1(4) & 6.1(2) & 0.46 & 0.24(6) & 0.3(1)  & -0.1(1) & 0.9 \\
 190 & 10.130(3) & 0.119(3) & 10.1(4) & 14.(2) & 9.3(9) & 3.0(4) & 5.1(2) & 0.32 & 0.98(3) & 0.161(6)  & -0.005(7) & 1.0 \\
 191 & 10.131(3) & 0.127(3) & 10.0(4) & 26.(3) & 7.0(6) & 5.2(5) & 4.8(1) & 0.29 & 0.77(4) & 0.16(1)  & -0.05(1) & 0.9 \\
 192 & 10.243(4) & 0.164(4) & 11.7(3) & 41.(4) & 3.(2) & 6.0(4) & 6.8(2) & 0.45 & 0.3(1) & 0.13(5)  & 0.08(5) & 0.9 \\

\enddata

\tablecomments{(1) Source Number from Table \ref{obstable}. (2) Source
  velocity with respect to the LSR. (3) Velocity dispersion of the
  NH$_3$ (4) Kinetic temperature. (5) Column Density of NH$_3$
  assuming $\eta_f=1$. (6) Column Density of C$_2$S. (7) Total opacity
  in the NH$_3$(1,1) line.  (8) Excitation Temperature for the
  NH$_3$. For lines-of-sight on which the opacity and excitation
  temperature cannot be separately determined, the product
  $\tau_{1}(T_{x}-T_{bg})$ is quoted, spanning both columns.  (9)
  Filling fraction for the case where $T_{K}=T_{x}$. (10) Peak
  temperature of C$_2$S. (11) Velocity dispersion of the of C$_2$S
  line. (12) Velocity offset of the C$_2$S line with respect to the
  NH$_3$ complex. (13) reduced $\chi^2$ for the fit (since data are
  not completely independent, this is only a goodness-of-fit parameter).}

\tablenotetext{a}{This property is affected by the overall amplitude
  calibration which is subject to an additional $\sim 5\%$ uncertainty. }
\tablenotetext{b}{The spectrum shows evidence for multiple components
in one of the spectral lines that {\it cannot} be resolved uniquely by
multicomponent fitting.}
\end{deluxetable}

%\clearpage
%\end{landscape}

\end{document}